\documentclass[varenna]{cimento}
\usepackage{epsf}
\usepackage{latexsym}
\begin{document}   

\newcommand{\tbox}[1]{\mbox{\tiny #1}}
\newcommand{\mboxs}[1]{\mbox{\small #1}} 
\newcommand{\mbf}[1]{{\mathbf #1}}
		    
\newcommand{\mpg}[2]{\begin{minipage}[t]{#1cm}{#2}\end{minipage}}	
\newcommand{\mpb}[2]{\begin{minipage}[b]{#1cm}{#2}\end{minipage}}
\newcommand{\mpc}[2]{\begin{minipage}[c]{#1cm}{#2}\end{minipage}}

\newcommand{\itm}{\\ $\bullet \ $}

\title{Chaos, Dissipation and Quantal Brownian Motion \\  
\normalsize (Lecture notes of course in the international school 
of physics {\em Enrico Fermi}, \\  
Session CXLIII {\em New Directions in Quantum Chaos}, Varenna Italy, July 1999)}

\shorttitle{Chaos, Dissipation and Quantal Brownian Motion}

\author{Doron Cohen}  
 
\institute{Department of Physics, Harvard University}  

\maketitle


\vspace*{-3cm}

\begin{abstract} 
Quantum dissipation, the theory of energy spreading 
and quantal Brownian motion are considered.  
In the {\em first part} of these lecture-notes we discuss 
the classical theory of a particle that interacts with 
chaotic degrees of freedom:    
\itm  The Sudden and the Adiabatic approximations; 
\itm  The route to stochastic behavior; 
\itm  The fluctuation-dissipation relation; 
\itm  And the application to the `piston' example. 

In the {\em second part} of these lecture-notes we discuss the 
restricted problem of classical particle that interacts with 
quantal (chaotic) degrees of freedom:  
\itm  Limitations on quantal-classical correspondence; 
\itm  The perturbative core-tail spreading profile; 
\itm  Linear response theory versus Fermi-golden-rule picture; 
\itm  Random-matrix-theory considerations; 
\itm  And the quantal Sudden and Adiabatic approximations.
  
In the {\em third part} of these lecture-notes we discuss the
problem of quantal particle that interacts with an effective 
harmonic bath:  
\itm  Classical Brownian motion; 
\itm  The ZCL model and the DLD model; 
\itm  The white noise approximation;  
\itm  The reduced propagator and master-equation formulation; 
\itm  And the two mechanisms for dephasing. 

We conclude with explaining the main open question 
in the theory of quantum dissipation and quantal Brownian motion. 
That question concerns the problem of quantal particle that interacts 
with quantal (chaotic) degrees of freedom. 
\end{abstract}

\section{Definition of the problem}

We are interested in the reduced dynamics 
of a slow degree of freedom $(\mbf{x},\mbf{p})$ 
that interacts with a `bath'. The Hamiltonian is 
\begin{eqnarray} 
{\cal H} \ = \ {\cal H}_0(\mbf{x},\mbf{p})
+{\cal H}_{\tbox{env}}(\mbf{x},Q_{\alpha},P_{\alpha}) 
\end{eqnarray}
In Fig.1 we display a list of our main assumptions 
and an illustration of our leading example.    
In this example the slow degree of 
freedom is the `piston', and the `bath' consist of one 
gas particle. 

\begin{figure}[h]
{\em Assumptions:} \\
$\bullet$ \ \  
${\cal H}_{\tbox{env}}(Q,P;\mbf{x})$ with $\mbf{x}{=}\mbox{const}$ 
generates classically chaotic dynamics. \\
$\bullet$ \ \  
${\cal H}_0(\mbf{x},\mbf{p})=\mbf{p}^2/(2m)$ 
describes a free particle. (optional is linear driving). \\ 
$\bullet$ \ \  
Initially the bath is characterized either by 
an energy $E$ or by a temperature $T$. \\
$\bullet$ \ \  
In a classical sense $\mbf{x}(t)$ is a slow degree of freedom. \\
\begin{center}
\leavevmode
\epsfysize=2.2in 
\epsffile{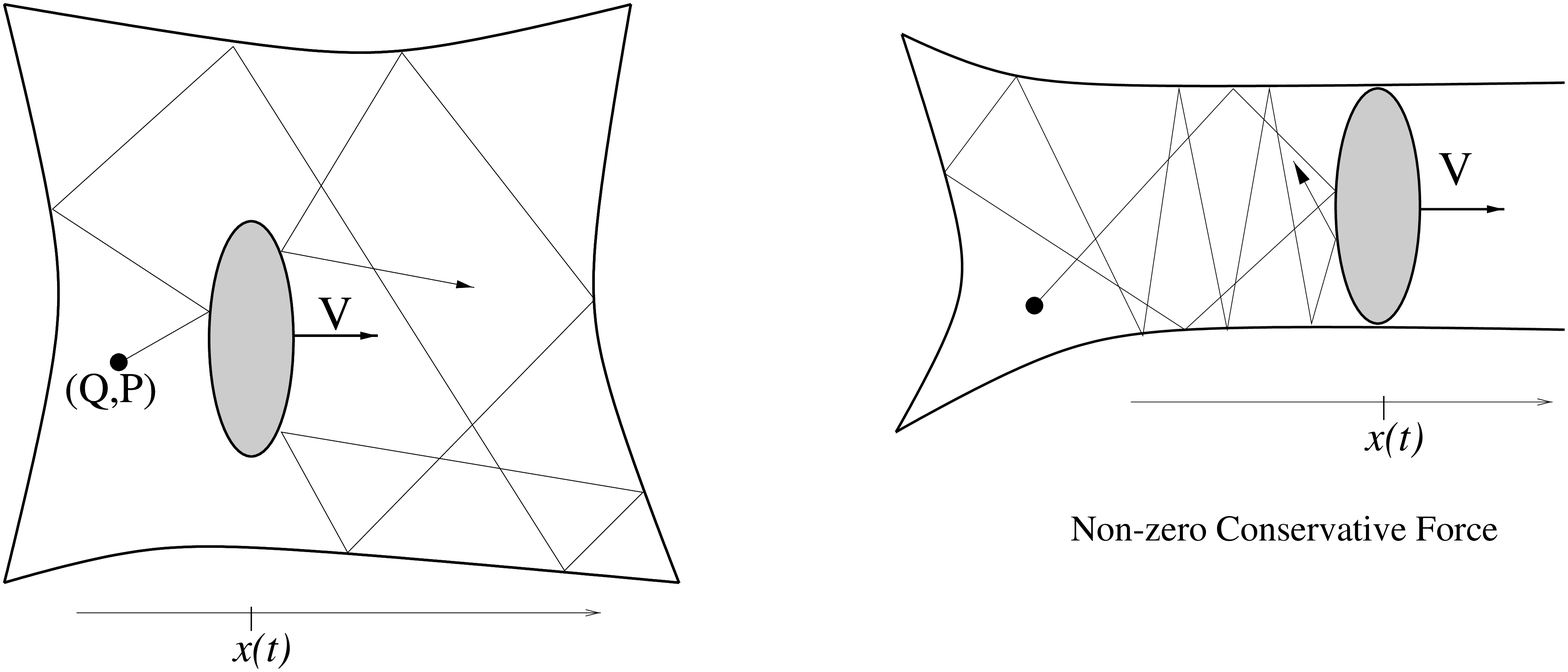}
\end{center}
\caption{\protect\footnotesize 
List of main assumptions and an 
illustration of the `piston' example. } 
\end{figure}

Classically the reduced dynamics of the 
slow degree of freedom is described 
by the Langevin equation
\begin{eqnarray} 
m\ddot{\mbf{x}} \ = \ {\cal F}(t)
\end{eqnarray}
where ${\cal F}(t)$ is a stochastic force. The average 
force on the particle is 
\begin{eqnarray} 
\langle {\cal F}(t) \rangle \ = \ 
F(\mbf{x}) - \mu \dot{\mbf{x}}
\end{eqnarray}
Without loss of generality, just for the sake of 
simplicity, we assume from now on that the 
velocity-independent force $F(\mbf{x})$ is 
equal to zero (see Fig.1). Thus there is 
no {\em reversible} energy change due to conservative work.     
However, there is still systematic {\em irreversible} change of  
energy due to the friction force. The energy 
dissipation rate is:  
\begin{eqnarray} 
\frac{d}{dt} \left\langle \ 
\frac{1}{2}m\dot{\mbf{x}}^2 \right\rangle 
\ \approx \ - \mu \dot{\mbf{x}}^2 \ \ \ \ \ 
\mbox{for non-thermal motion.}
\end{eqnarray}
The fluctuating component of the stochastic force 
is like white noise, and it is characterized by its 
intensity $\nu$. The dissipation constant $\mu$ is 
related to the noise intensity $\nu$ via a universal 
Fluctuation-Dissipation (FD) relation 
\begin{eqnarray} 
\mu \ = \ \frac{\nu}{2k_{\tbox{B}}T}
\end{eqnarray}

We are interested in developing a corresponding 
{\em quantum mechanical theory of dissipation and 
quantal Brownian motion}. `Dissipation' means from 
now on that energy is absorbed by the bath degrees 
of freedom due to the time dependence of $\mbf{x}$. 
The dissipation coefficient is $\mu$.  
`Brownian motion' refers from now on to the reduced 
dynamics of the $(\mbf{x},\mbf{p})$ degree of freedom. 
We shall use  the term `piston' rather than 
`Brownian particle' if the motion is constrained to 
one dimension.  Note that the term `piston' is not used 
in its literal sense.

\section{Restricted versions of the problem} 
  
The quantum mechanical treatment of the general problem is 
extremely complicated. Therefore it is a good idea to 
analyze restricted versions of the general problem. 
Treating $\mbf{x}(t)$ as a classical 
degree of freedom, we can consider the time dependent Hamiltonian 
\begin{eqnarray} \label{e6}
{\cal H} \ = \ {\cal H}_{\tbox{env}}(Q,P;\mbf{x}(t))  
\end{eqnarray}
For simplicity we can further assume that $\mbf{x}(t)$ 
describes a motion with a constant velocity $\dot{\mbf{x}}=V$.
Consequently we can treat $x$ in (\ref{e6}) as a 
one-degree-of-freedom variable. This restricted problem is 
the main issue of the following lectures (parts 1 and 2).

In consistency with the terminology that has been introduced 
at the end of the previous section, we shall refer to the 
restricted problem defined above as 
{\em `the problem of quantum dissipation'}. 
Obviously, in other physical examples $x$ does not have 
to be the position of a particle. It can be any controlling 
parameter that appears in the Hamiltonian, e.g. electric field.   
Of particular interest is the case where $x$ is the 
magnetic flux via a ring. The velocity $V=\dot{x}$ has 
then the meaning of electro-motive-force.  Let us 
assume that the ring contains one charged particle $(Q,P)$ 
that performs diffusive motion. The charged particle 
gains kinetic energy, and the dissipation coefficient $\mu$ 
is just the conductivity of the ring. 
Note that in actual circumstances the charged-particle 
is an electron, and its (increasing) kinetic energy 
is eventually transfered to the vibrational modes (phonons) 
of the ring, leading to {\em Joule heating}.  
The latter process is `on top' of the generic dissipation 
problem that we are going to analyze.

We come back to the Brownian particle / `piston' example, 
and shift the focus from the `bath' to the reduced-dynamics 
of the $(\mbf{x},\mbf{p})$ degree-of-freedom. We can study 
the effect of the fluctuating force by considering 
an Hamiltonian of the type 
\begin{eqnarray} \label{e7}
{\cal H} \ = \ \frac{\mbf{p}^2}{2m} + 
{\cal U}(\mbf{x},t)  
\end{eqnarray}
where ${\cal U}(\mbf{x},t)$ is an effective  
stochastic potential the mimics the noisy character 
of the environment. Obviously, with such Hamiltonian 
we cannot mimic the effect of dissipation. In order 
to have dissipation the interaction should be with 
dynamical  degrees of freedom. We can introduce an 
effective harmonic-bath as follows: 
\begin{eqnarray} \label{e8}
\hspace*{-2cm} 
{\cal H} \ = \ \frac{\mbf{p}^2}{2m}  
\ - \ \sum_{\alpha} c_{\alpha} Q_{\alpha} 
u(\mbf{x}{-}\mbf{x}_{\alpha})
\ + \ \sum_{\alpha}\left
(\frac{P_{\alpha}^2}{2m_{\alpha}}
+\frac{1}{2} m \omega_{\alpha}^2 Q_{\alpha}^2\right)
\end{eqnarray}
The effective-noise model (\ref{e7}) as well as the 
effective harmonic-bath model (\ref{e8}) can be treated 
analytically using the Feynman-Vernon (FV) formalism. 
{\em The reduced dynamics of the particle} is obtained after 
averaging over realizations of the stochastic 
potential (in case of (\ref{e7})), or by elimination
of the environmental degrees of freedom
(in case of (\ref{e8})). It leads to a unified 
description of {\em diffusion localization and dissipation} (DLD). 
It is found indeed that the 
formal solution of (\ref{e8}) reduces to that 
of (\ref{e7}) if dissipation effect is neglected.  
We shall refer to (\ref{e8}), and more generally 
to its formal solution, as the `DLD model'.   
It is possible to introduce an effective 
RMT-bath instead of an effective harmonic bath, 
but then there is no general analytical solution.  
It has been demonstrated however that at high temperatures 
an approximate treatment of the effective RMT-bath 
coincides with the exact (high temperature) solution 
of the DLD model. 

\section{`History' of the problem 
{\rm (possibly biased point of view \cite{rmrk})} }

There is a {\em well established classical theory for 
dissipation and Brownian motion}. In the {\em pre-chaos}  
literature (${\sim}1980$) the dissipation constant for  
a moving `piston' is given by the `wall formula'. 
See \cite{wall,koon} and followers. 
A more general point of view, which is 
discussed in the {\bf first part} of these lectures 
(Sections 4-11),  
has been adopted in the {\em post-chaos} literature (${\sim}1990$). 
There, the emphasis is on relating the 
dissipation constant to the  intensity of fluctuations  
\cite{ott,wilk,jar,berry}. 

Various methods such as `linear response', 
`Kubo-Greenwood formalism', and `multiple scale analysis' 
have been used \cite{koon,wilk,berry}
in order to make a quantum mechanical derivation  
of the FD relation. All these methods 
are essentially equivalent to a naive 
application of the Fermi-Golden-Rule (FGR) picture. 
The validity of the {\em naive} FGR result has been 
challenged in by Wilkinson and Austin \cite{WA}. 
They came up with a surprising conclusion that we would 
like to paraphrase as follows: 
A proper FGR picture, supplemented by an innocent-looking 
RMT assumption, leads to a {\em modified} FGR result;  
In the classical limit the modified FGR result disagrees 
with the FD relation and leads to violation 
of the quantal-classical correspondence principle. 
Obviously this conclusion should be regarded as 
a provocation for constructing {\em a proper theory 
for Quantum Dissipation} \cite{crs,frc}. This is the issue of the 
{\bf second part} of these lectures (Sections 12-24). 
The theory is strongly related to some studies of parametric 
dynamics \cite{wigner,flamb,casati,felix2} 
and wavepacket dynamics \cite{qkr,kottos,wbr}.

Most of the literature that comes under the heading 
`Quantum Dissipation' is concerned with the more general 
problem where $x$ is a dynamical variable \cite{textbook}.  
A direct handling of the `bath' degrees of freedom is 
usually avoided. As an exception see for example \cite{kolovsky}. 
It is common to adopt an `effective-bath' approach. 
See \cite{FV,ZG,MS} and followers.  
Specific discussion of quantal Brownian motion has been 
introduced in \cite{CL} and later in \cite{dld,qbm}. 
It leads to the DLD model, that unifies the treatment of 
diffusion localization and dissipation. 
This is  the issue of the {\bf third part} of these lectures
(Sections 25-31). 
The high temperature version of the DLD model is 
obtained also by considering coupling to an effective 
RMT-Bath \cite{rmt}.

\section{Fluctuations: intensity and correlation time }

We consider the Hamiltonian ${\cal H}(Q,P;x)$ with 
$x{=}\mbox{const}$. The phase space volume which is  
enclosed by the energy surface ${\cal H}(Q,P;x)=E$ 
will be denoted by $\Omega(E;x)$. For the classical 
density of states we shall use the notation 
$g(E)= \partial_{\tbox{E}}\Omega(E)$. We define
\begin{eqnarray}
{\cal F}(t) \ = \  
-\frac{\partial {\cal H}}{\partial x}(Q(t),P(t);x)
\end{eqnarray}
The conservative force 
$F(x)=\langle {\cal F}(t) \rangle_{\tbox{E}}$ 
is obtained by performing microcanonical 
averaging over $(Q,P)$. It is  
equal to zero if and only if 
$\Omega(E;x)$ is independent of~$x$.  
For simplicity we assume from now on that 
this is indeed the case.   
Now we define a correlation function 
\begin{eqnarray}
C_{\tbox{E}}(\tau) \ = \ \langle 
{\cal F}(t) {\cal F}(t{+}\tau) \rangle_{\tbox{E}}
\end{eqnarray}
The fluctuating force is characterized by an intensity 
\begin{eqnarray}
\nu_{\tbox{E}} \ = \ 
\int_{-\infty}^{\infty} C_{\tbox{E}}(\tau) d\tau
\end{eqnarray}
and by a correlation time $\tau_{\tbox{cl}}$.
The power spectrum of the fluctuations  
$\tilde{C}_{\tbox{E}}(\omega)$ is the 
Fourier transform of $C_{\tbox{E}}(\tau)$. 
For chaotic bath the stochastic force is 
like white noise.   
We always have $\tau_{\tbox{cl}} \le t_{\tbox{erg}}$, 
where  $t_{\tbox{erg}}$ is the ergodic time. 
In the general discussion we shall not distinguish 
between the two time scales. However, in specific 
examples the distinction is meaningful. For example, 
in case of the `piston' example $\tau_{\tbox{cl}}$ 
is the collision time with the walls of the piston, 
while $t_{\tbox{erg}}$ is of the order of the 
ballistic time~$\tau_{\tbox{bl}}$.

\section{Fluctuations: time dependent Hamiltonian}
\label{s5}

We consider from now on the time-dependent Hamiltonian 
${\cal H}(Q,P;x(t))$ with 
constant non-zero velocity $\dot{x}=V$.
We define 
\begin{eqnarray}
{\cal F}(t) \ = \  
-\frac{\partial {\cal H}}{\partial x}(Q(t),P(t);x(t))
\end{eqnarray}
The statistical properties of the 
fluctuating force ${\cal F}(t)$ are 
expected to be slightly different from the 
$V{=}0$ case. The average $\langle {\cal F}(t) \rangle$ 
is no longer expected to be zero. 
Rather, we expect to have  
$\langle {\cal F}(t) \rangle = -\mu V$. 
This implies that the correlator 
$\langle {\cal F}(t_1) {\cal F}(t_2) \rangle$ 
acquires an offset $(\mu V)^2$. We shall 
argue that if the velocity $V$ is small enough 
then the `offset correction' can be ignored for 
a relatively long time which will be denoted 
by $t_{\tbox{frc}}$. Obviously it is essential 
to have
\begin{eqnarray}   
\tau_{\tbox{cl}} \ \ll \ t_{\tbox{frc}}
\ \ \ \ \ \mbox{[non-trivial slowness condition].}
\end{eqnarray}
There is another possible reason for the correlator 
$\langle {\cal F}(t_1) {\cal F}(t_2) \rangle$ 
to be different from  $C_{\tbox{E}}(t_2{-}t_1)$. 
Loss of correlation may be either due to the 
dynamics of $(Q(t),P(t))$ or else due to the 
parametric change of $x(t)$.  We can define a 
parametric correlation scale $\delta x_c^{\tbox{cl}}$. 
For the piston example it is just the penetration 
distance into the piston upon collision (the effective 
`thickness' of the wall).  The associated parametric 
correlation time is $\tau_c^{\tbox{cl}}=\delta x_c^{\tbox{cl}}/V$.  
We assume that   
\begin{eqnarray}   
\tau_{\tbox{cl}} \ \ll \ \tau_c^{\tbox{cl}}
\ \ \ \ \ \mbox{[trivial slowness condition].}
\end{eqnarray}
meaning that loss of correlations is predominantly 
determined by the chaotic nature of the 
dynamics rather than by the (slow) parametric 
change of the Hamiltonian. 
For the piston example application of 
the above requirements leads to the obvious 
condition $V \ll v_{\tbox{E}}$, where $v_{\tbox{E}}$ 
is the velocity of the gas particle.  

\section{Actual, Parametric and Reduced energy changes}

For the time dependent Hamiltonian ${\cal H}(Q,P;x(t))$ 
energy is not a constant of the motion. 
Changes in the actual energy ${\cal E}(t)$ reflect 
`real' dynamical changes as well as parametric changes. 
Therefore it is useful to introduce the following definitions: 
\begin{eqnarray}
{\cal E}(t) \ \ = \ {\cal H}(Q(t),P(t);x(t)) \\
{\cal E}'(t) \ = \ {\cal H}(Q(t),P(t);x(0)) 
\end{eqnarray}
The actual energy change can be calculated as follows:
\begin{eqnarray}
\frac{d{\cal E}}{dt} \ = \ 
\frac{\partial {\cal H}}{\partial t} \ = \ 
-{\cal F}(t) \cdot V  \\
\delta{\cal E} \ = \ {\cal E}(t)-{\cal E}(0) \ = \ 
- V \int_0^t {\cal F}(t') dt' 
\end{eqnarray}
The actual energy change $\delta{\cal E}$ can be 
viewed as a sum of parametric-energy-change~$\delta{\cal E}_o$, 
and reduced-energy-change~$\delta{\cal E}'$.
\begin{eqnarray}
\delta{\cal E} \ \ & = & \
{\cal H}(Q(t),P(t);x(t)) - {\cal H}(Q(0),P(0);x(0)) 
\\
\delta{\cal E}_o  \ & = & \
{\cal H}(Q(t),P(t);x(t)) - {\cal H}(Q(t),P(t);x(0)) 
\\
\delta{\cal E}' \ & = & \ 
{\cal H}(Q(t),P(t);x(0)) - {\cal H}(Q(0),P(0);x(0))
\end{eqnarray}
The reduced energy change $\delta{\cal E}'$ reflects 
the deviation of $(Q,P)$ from the original energy surface. 
It can be calculated as follows:
\begin{eqnarray}
\delta{\cal E}' \ = \  
V \cdot \left[{\cal F}(t)\times t 
\ - \ \int_0^t {\cal F}(t')dt' \right]
\end{eqnarray}
On the other hand, the actual energy change 
$\delta{\cal E}$ reflects the deviation of 
$(Q,P)$ from the instantaneous energy surface.

\section{The Sudden and the Adiabatic approximations}

\begin{figure}
\begin{center}
\leavevmode
\epsfysize=1.7in \epsffile{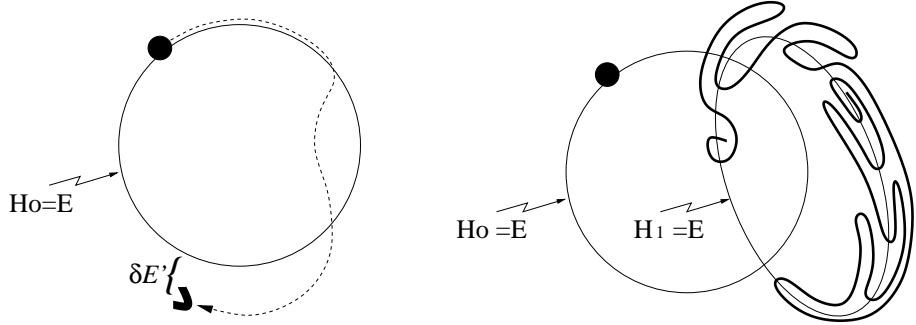}
\end{center}
\caption{\protect\footnotesize 
Schematic illustration of the dynamics.
An initially localized distribution is launched in phase space.  
For a limited time (left plot) it travels upon the 
initial energy surface. But then it departs from it. 
After much longer times (right plot) the evolving 
distribution is concentrated across an  
instantaneous energy surface. } 
\label{f_evolving}
\end{figure}

By inspection of the expressions for the reduced 
energy change we arrive at the conclusion that for 
short times we have the so called `sudden approximation':
\begin{eqnarray}
\delta{\cal E}' \ \approx \ 0 
\ \ \ \ \ \mbox{for $t\ll \tau_{\tbox{cl}}$}
\end{eqnarray}
By inspection of the expression for the actual energy 
change we arrive at the conclusion that for 
longer times we have the so called `adiabatic approximation':
\begin{eqnarray}
\delta{\cal E} \ \sim \ 0 
\ \ \ \ \ \mbox{for $t\ll t_{\tbox{frc}}$}
\end{eqnarray}
The time evolution of an initially localized 
phase-space distribution $\rho_0(Q,P)$ is illustrated 
in Fig.\ref{f_evolving}.  
For short times we have in general non-stationary time evolution:
\begin{eqnarray} 
{\cal U}(t) \ \rho_0(Q,P) \ \ne \ \rho_0(Q,P)
\ \ \ \ \ \ \ \ \ \ \ \ \ \ \ \  
\mbox{for $t\ll\tau_{\tbox{cl}}$}
\end{eqnarray}
Here ${\cal U}(t)$ is the classical propagator of 
phase space points. However, if we operate  
with the same ${\cal U}(t)$ on a microcanonical 
distribution, then
\begin{eqnarray} 
{\cal U}(t) \ \rho_{\tbox{E,x(0)}}(Q,P) 
\ \approx \ \rho_{\tbox{E,x(0)}}(Q,P)
\ \ \ \ \ \ \ \mbox{for $t\ll\tau_{\tbox{cl}}$}
\end{eqnarray}
It is as if the microcanonical state does not have 
the time to adjust itself to the changing Hamiltonian.  
The sudden approximation implies that for 
short times ${\cal U}(t)$ can be replaced by 
{\em unity} if it operates on an initial microcanonical 
state. For longer times we have 
\begin{eqnarray} 
{\cal U}(t) \ \rho_0(Q,P) 
\ \sim \ \rho_{\tbox{E,x(t)}}(Q,P) \ \ \ \ \ \ \ 
\mbox{for $t_{\tbox{erg}} \ll t \ll t_{\tbox{frc}}$}
\end{eqnarray} 
Here there is enough time for the evolving distribution 
to adjust itself to the changing Hamiltonian. If we 
start with a microcanonical state, then the above 
similarity will hold for any $t \ll t_{\tbox{frc}}$. 
The adiabatic approximation becomes 
worse and worse as time elapses due to the 
transverse spreading across the energy surface.
We shall see that $t_{\tbox{frc}}$ is the 
breaktime for the adiabatic approximation.

\section{Ballistic and Diffusive energy spreading}

We recall the formula for the actual energy change 
\begin{eqnarray}
{\cal E}(t)-{\cal E}(0) \ = \ 
- V \int_0^t {\cal F}(t') dt' 
\end{eqnarray} 
We Assume an initial microcanonical 
preparation which is characterized 
by an energy~$E$. The energy spreading 
after time $t$ is: 
\begin{eqnarray}
\langle ({\cal E}(t)- {\cal E}(0))^2 \rangle \ = \ 
V^2 \int_0^t \int_0^t 
\langle {\cal F}(t_1) {\cal F}(t_2) \rangle
\ dt_1dt_2
\end{eqnarray} 
Now we make the following approximation
\begin{eqnarray}
\langle {\cal F}(t_1) {\cal F}(t_2) \rangle
\ \approx \ C_{\tbox{E}}(\tau)
\ \ \ \ \ \ \ \ \ \ 
\mbox{applicable if $\ t\ll t_{\tbox{frc}}$}
\end{eqnarray} 
The validity of this approximation is restricted 
by the condition $t\ll t_{\tbox{frc}}$ which will be 
discussed later. The energy spread is 
\begin{eqnarray} \label{e25}
\langle ({\cal E}(t)-{\cal E}(0))^2 \rangle 
\ \approx \ C_{\tbox{E}}(0) \cdot (Vt)^2
\ \ \ \ \ \ \ \ \ \ 
& \mbox{for $t \ll \tau_{\tbox{cl}}$} \\
\langle ({\cal E}(t)-{\cal E}(0))^2 \rangle 
\ \approx \ 2D_{\tbox{E}} \ t
\ \ \ \ \ \ \ \ \ \ \ \ \ \ \ \ \ \  
& \mbox{for $\tau_{\tbox{cl}} \ll t \ll t_{\tbox{frc}}$} 
\end{eqnarray} 
The ballistic spreading on short time scales just 
reflects the parametric change of the energy surfaces. 
The diffusive spreading on longer times reflects 
the deviation from the adiabatic approximation. 
The diffusion coefficient is 
\begin{eqnarray}
D_{\tbox{E}} \ \ = \ \ 
\frac{1}{2}V^2 \int_{-t}^t C_{\tbox{E}}(\tau)d\tau
\ \ \rightarrow \ \
\frac{1}{2} \ \nu_{\tbox{E}} \ V^2
\end{eqnarray}

\section{Energy spreading and dissipation}

It is possible to argue that the energy distribution 
$\rho(E)$ obeys the following diffusion equation: 
\begin{eqnarray}
\frac{\partial \rho}{\partial t} \ = \ 
\frac{\partial}{\partial E}
\left[g(E)D_{\tbox{E}} \ \frac{\partial}{\partial E}
\left(\frac{1}{g(E)}\rho\right)\right]
\ \ \ \ \ \ \ \ \mbox{for $t \gg t_{\tbox{erg}}$}
\end{eqnarray} 
If we transform to the proper phase space variable 
$n=\Omega(E)$, this diffusion equation gets the standard form.  
The average energy is calculated via
\begin{eqnarray}
\langle {\cal E}(t) \rangle = \int_0^{\infty} E \ \rho(E)dE
\end{eqnarray} 
and consequently
\begin{eqnarray}
\frac{d}{dt}\langle {\cal E} \rangle 
\ = \ - \int_0^{\infty} dE \ g(E) D_{\tbox{E}}  
\ \frac{\partial}{\partial E}
\left(\frac{\rho(E)}{g(E)}\right) 
\ \ \equiv \ \ \mu V^2
\end{eqnarray} 
Using the expression for  $D_{\tbox{E}}$ We obtain the 
following FD relation
\begin{eqnarray}
\mu_{\tbox{E}} \ = \ \frac{1}{2} \frac{1}{g(E)} 
\frac{\partial}{\partial E}
(g(E) \nu_{\tbox{E}})
\end{eqnarray} 
The actual value of $\mu$ is obtained by averaging 
$\mu_{\tbox{E}}$ according to the distribution $\rho(E)$. 
The standard version of the FD relation 
$\mu=\nu/(2k_{\tbox{B}} T)$ is obtained if $\rho(E)$ 
is of the canonical type.  

\section{Application to the `piston' example}

\begin{figure}
\begin{center}
\leavevmode 
\epsfysize=6.0in 
\epsffile{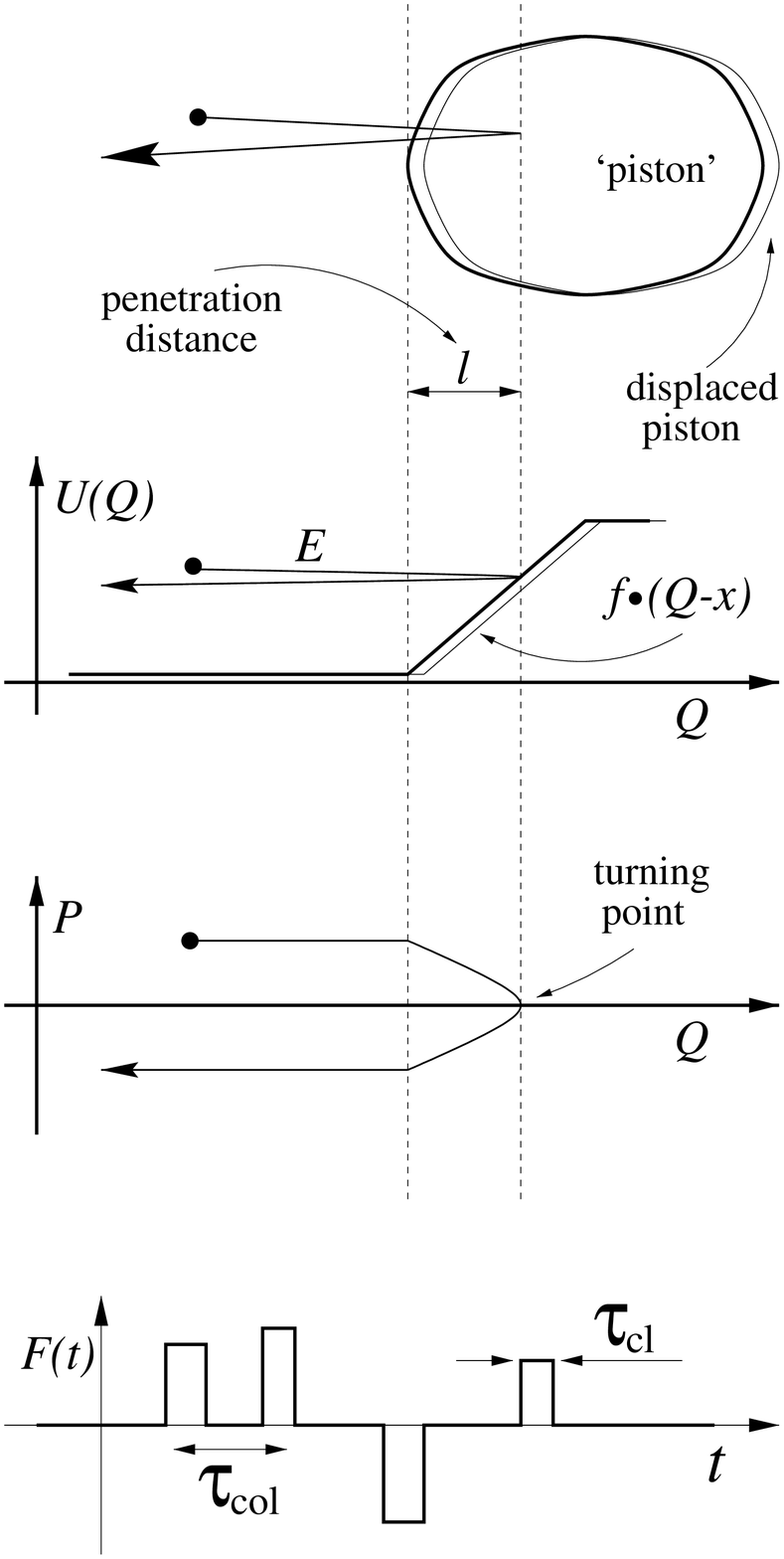}
\end{center}
$\bullet$ \ \ \ 
Assume that the velocities 
$\ |v_{\tbox{col}}| \ \sim \ v_{\tbox{E}} \ $ 
are uncorrelated. \\
$\bullet$ \ \ \ 
The average time 
between collisions is 
$\tau_{\tbox{col}} \ = \ 
(\mbox{\small Volume}/\mbox{\small Area}) \ / \ v_{\tbox{E}}$. \\
$\bullet$ \ \ \ 
For $V=0$ we have $\langle {\cal F}(t) \rangle = 0$ and 
$\nu_{\tbox{E}} = (2\mathsf{m}v_{\tbox{E}})^2 \cdot (1/\tau_{\tbox{col}})$. \\
$\bullet$ \ \ \ 
For $V\ne 0$ we have 
$\langle {\cal F}(t) \rangle  =  -\mu_{\tbox{E}}V$ where
$\mu_{\tbox{E}}  =  2\mathsf{m} \cdot (1/\tau_{\tbox{col}})$ \\
$\bullet$ \ \ \ 
The phase space volume is 
$\Omega(E) = \mbox{\small Volume} \cdot (\mathsf{m}v_{\tbox{E}})^d$. \\
\caption{\protect\footnotesize 
Illustration of one collision with the piston, 
and a list of steps in the analysis of a series 
of such collisions. The {\small Area} of the moving faces,  
the {\small Volume} of the cavity, the mass $\mathsf{m}$ of the 
gas-particle, and its kinetic energy $E$ should be specified.  
Geometric $d$-dependent prefactors have been dropped. 
See \cite{frc} for exact calculations. 
} 
\label{f_col}
\end{figure}

It is instructive to apply the FD relation for 
the `piston' example.  The bath degrees of freedom is 
a gas particle whose mass is $\mathsf{m}$. The faces of the 
piston are characterized by their total Area.
The piston is moving with a constant velocity $V$ 
inside a $d$-dimensional cavity. For simplicity we treat the 
collisions as `one-dimensional' and omit $d$ dependent 
pre-factors \cite{frc}.  
An illustration of one collision with the piston 
is presented in Fig.\ref{f_col}. Each collision can 
be either from the left or from the right side. 
The resultant stochastic force which is experienced by the 
piston is a sum over short impulses:    
\begin{eqnarray}
{\cal F}(t) \ = \ \sum_{\tbox{col}}
[ 2\mathsf{m}(v_{\tbox{col}} - V) ] \ \delta(t-t_{\tbox{col}})  
\end{eqnarray} 
The duration of each impulse is equal to the penetration-time 
upon a collision with the (soft) faces of the piston. 
If successive collisions with the piston are uncorrelated, 
then the correlation time $\tau_{\tbox{cl}}$ is equal to 
the average duration of an impulse.  
The main steps in the analysis of the multi-collision 
process are summarized in Fig.\ref{f_col} as well.  
It is easily verified that the FD relation between 
$\mu_{\tbox{E}}$ and $\nu_{\tbox{E}}$ is satisfied. 
Note however that a proper application of kinetic theory is 
essential in order to obtain the correct geometrical 
factors involved \cite{frc}.

The FD relation is a very powerful tool . This becomes 
most evident once we consider a variation of the above example: 
If successive collisions with the piston are correlated, for example 
due to bouncing behavior, then it is still a relatively 
easy task to estimate $\nu_{\tbox{E}}$ for the $V=0$ case, and 
then to obtain $\mu_{\tbox{E}}$ via the FD relation. 
On the other hand, a direct evaluation of $\mu_{\tbox{E}}$ 
using kinetic considerations is extremely difficult, because  
in calculating  $\langle{\cal F}(t)\rangle$ it is essential to 
take into account the correlations between successive collisions.

\section{The route to stochastic behavior}

\begin{figure}
\begin{center}
\leavevmode 
\epsfysize=2.5in 
\epsffile{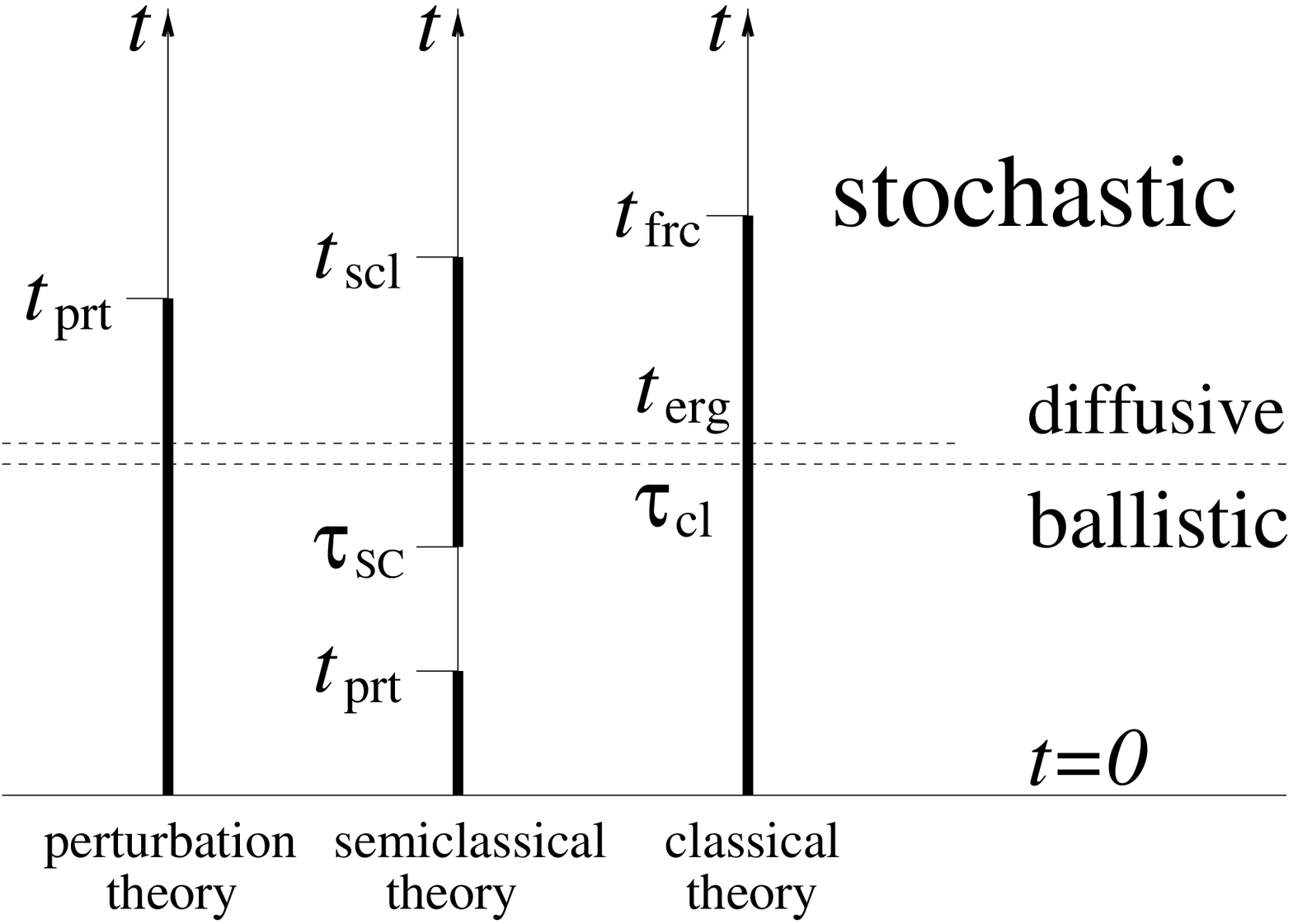} 
\vspace*{5mm}
\frame{\ \ \ \ \ \ \mpg{9}{ \ \\
$t_{\tbox{frc}} \ = \ \mbox{breakdown of classical perturbation theory}$ \\
$t_{\tbox{prt}} \ = \ \mbox{breakdown of quantal perturbation theory}$ \\ 
$t_{\tbox{scl}} \ = \ \mbox{breakdown of semiclassical approximation}$ \\ 
$\tau_{\tbox{cl}} \ll t_{\tbox{frc}}$
$ \ \ \ \ \ \ \leadsto \ \ \ \ \ \ $
\mbox{classical definition of slowness} \\
$\tau_{\tbox{cl}} \ll t_{\tbox{prt}}$
$ \ \ \ \ \ \ \leadsto \ \ \ \ \ \ $
\mbox{quantal definition of slowness} \\
$\tau_{\tbox{cl}} \ll t_{\tbox{scl}}$
$ \ \ \ \ \ \ \leadsto \ \ \ \ \ \ $
\mbox{not restrictive condition} \\
$\tau_{\tbox{SC}} \ll \tau_{\tbox{cl}}$
$ \ \ \ \ \ \ \leadsto \ \ \ \ \ \ $
\mbox{quantal definition of fastness}
\ \\ }}
\end{center}
\caption{\protect\footnotesize 
Illustration of the various time scales involved in 
constructing either classical, semiclassical or 
perturbation theory of dissipation. 
We use the notation $\tau_{\tbox{SC}}=\delta x_{\tbox{SC}}/V$.  
The accompanying table summarizes the associated requirements for the 
applicability of each of those theories. Note that 
the quantum mechanical definitions of slowness and 
of fastness are not complementary. In any case, 
slowness in the classical sense is always assumed.} 
\label{f1}
\end{figure}

The derivation of the FD relation consists of two 
steps: The {\em first step} establishes the local 
diffusive behavior for short ($t\ll t_{\tbox{frc}}$) 
time scales, and $D_{\tbox{E}}$ is determined;
The {\em second step} establishes the global stochastic 
behavior on large ($t\gg t_{\tbox{erg}}$) time scales. 
The various time scales involved are illustrated 
in Fig.\ref{f1}. The classical breaktime $t_{\tbox{frc}}$ 
is defined as follows:
\begin{eqnarray}
t_{\tbox{frc}} \ = \ \nu/(\mu V)^2
\end{eqnarray} 
For $t>t_{\tbox{frc}}$ the systematic energy change 
$\langle \delta {\cal E} \rangle = \mu V^2t$ 
becomes larger than the energy spreading 
$(\langle \delta {\cal E}^2 \rangle)^{\tbox{1/2}} 
= (\nu V^2t)^{\tbox{1/2}}$,   
and the local analysis becomes meaningless. 
Therefore $t_{\tbox{frc}}$ can be regarded as the 
breakdown of the adiabatic approximation.

From the discussion above it follows that the validity 
of the classical derivation is restricted by the non-trivial 
slowness condition $\tau_{\tbox{cl}} \ll t_{\tbox{frc}}$. 
Alternatively, it is more useful to regard the non-trivial 
slowness condition using the point-of-view of Section 5. 
This point-of-view is further explained below, 
and can be easily generalized to the quantum-mechanical case.
For $V\ne 0$ the average of ${\cal F}(t)$ does not 
vanish, and therefore its correlator 
should include an `offset' term. Namely, 
$\langle {\cal F}(t_1) {\cal F}(t_2) \rangle 
\approx C_{\tbox{E}}(t_2{-}t_1) + (\mu V)^2$.  
The offset term can be neglected for 
a limited time $t \ll t_{\tbox{frc}}$ 
provided $(\mu V)^2 \ll C_{\tbox{E}}(0)$. 
The latter condition is equivalent to 
the non-trivial slowness condition  
$\tau_{\tbox{cl}} \ll t_{\tbox{frc}}$.

In the quantum-mechanical analysis we can use 
the same two-steps strategy in order to derive a 
corresponding FD relation.  
{\em We are going to concentrate on the first step}. 
It means that our objective is 
to establish a crossover from ballistic to diffusive behavior 
at the time $t\sim\tau_{\tbox{cl}}$.  The classical 
analysis in Section 8 is essentially `perturbation theory'. 
We can follow formally the same steps in the quantum-mechanical 
derivation, using Heisenberg picture.    
However, quantum-mechanical perturbation-theory 
is much more fragile than the corresponding classical theory, 
and we have typically $t_{\tbox{prt}} \ll t_{\tbox{frc}}$.  
Therefore, the quantum mechanical definition of slowness  
($\tau_{\tbox{cl}} \ll t_{\tbox{prt}}$) is much more 
restrictive than the classical requirement 
($\tau_{\tbox{cl}} \ll t_{\tbox{frc}}$).  
In the regime where perturbation theory fails we 
have to use a non-perturbative theory. In particular we 
are going to find the sufficient conditions for having 
detailed quantal-classical correspondence (QCC) using 
semiclassical considerations.

\section{The transition probability kernel}

In order to go smoothly from the classical theory 
to the quantum mechanical theory it is essential 
to use proper notations. From now on we 
shall use the variable $n = \Omega(E)$ instead of $E$. 
Given $x$, the energy surface that corresponds to
a phase-space volume $n$ will be denoted by:
\begin{eqnarray}
|n(x)\rangle \ = \ 
\{(Q,P) | \ \ {\cal H}(Q,P;x)=E_n \ \}
\end{eqnarray}
where $E_n$ is the corresponding energy. 
The microcanonical distribution which is 
supported by $|n(x)\rangle$ will be 
denoted by $\rho_{n,x}(Q,P)$. Upon quantization 
the variable $n$ becomes a level-index, and 
$\rho_{n,x}(Q,P)$ should be interpreted as 
the Wigner function that corresponds to the 
eigenstate $|n(x)\rangle$. With these definitions 
we can address the quantum mechanical theory and  
the classical theory simultaneously. We shall 
use from now on an admixture of classical 
and quantum-mechanical jargon. This should not 
cause any confusion.  

\begin{figure}
\begin{center}
\leavevmode 
\epsfysize=6.0in 
\epsffile{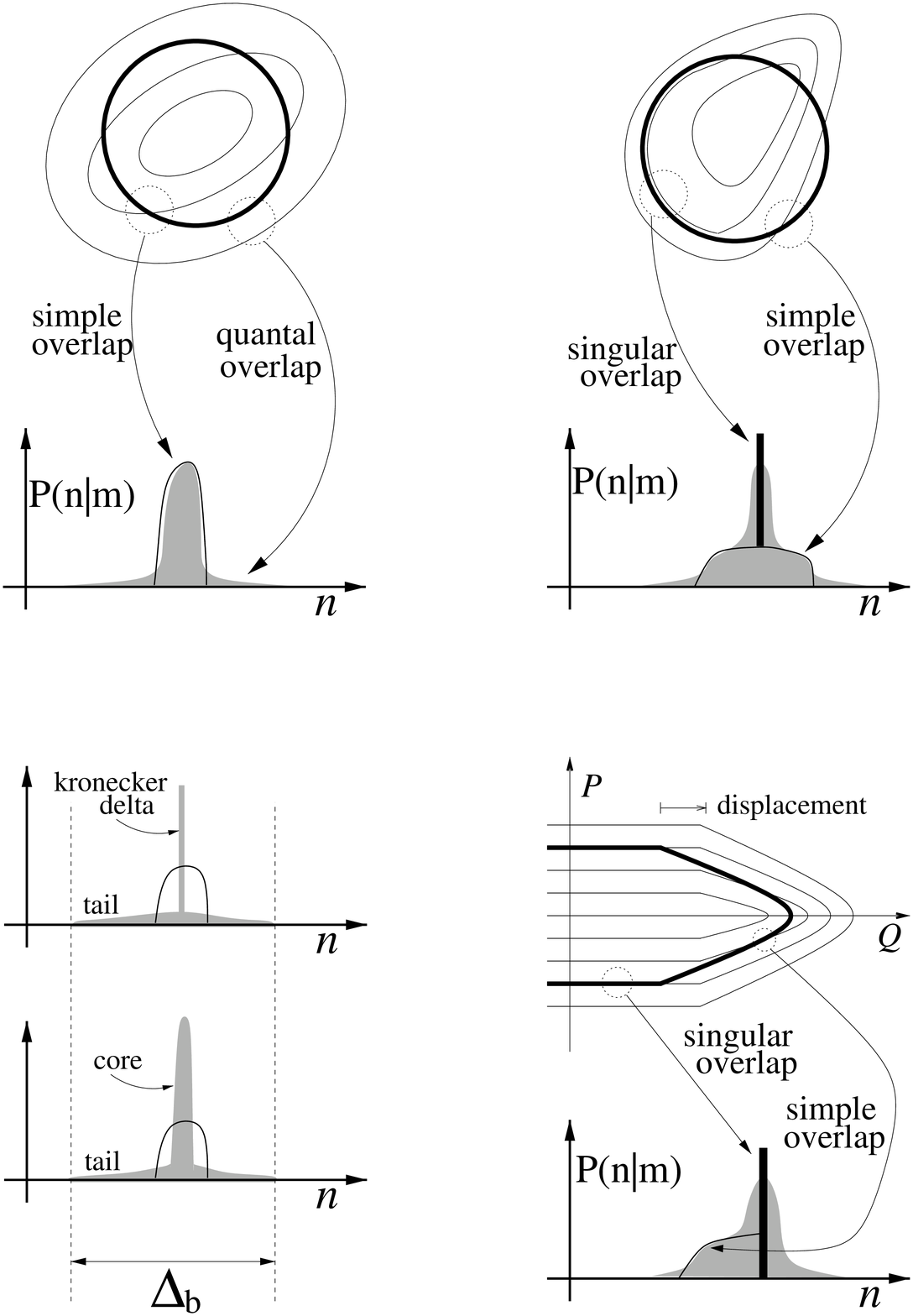}
\end{center}
\caption{\protect\footnotesize 
{\em Upper Left:} Phase space illustration of the 
initial and of the instantaneous set of parametric 
energy surfaces; Plot of the associated $P(n|m)$, 
where the classical behavior is indicate 
by the black lines, and the quantum-mechanical 
behavior is represented by the grey filling.   
Detailed QCC is assumed.  In the quantum-mechanical case  
classical sharp-cutoffs are being smeared.  
{\em Upper Right:} Illustration of a typical 
non-generic feature. In the quantum-mechanical case 
the classical delta-singularity is being smeared. 
{\em Lower Right:} The same non-generic feature 
manifests itself in the `piston' example. 
{\em Lower Left:} In the perturbative case there 
is no detailed QCC. The kernel is characterized by a core-tail 
structure. The tail is limited by the bandwidth 
of the coupling matrix-elements. If $\delta x$ is 
sufficiently small the core is just a Kronecker's delta.} 
\label{f_surfaces}
\end{figure}

The transition probability kernel $P_t(n|m)$ is the 
projection of the evolving state on the 
instantaneous set of energy surfaces. 
It is also possible to define a parametric kernel
$P(n|m)$. See illustrations in Fig.\ref{f_surfaces}. 
The definitions are:
\begin{eqnarray} \label{Ptnm}
P_t(n|m) \ & = & \ \mbox{trace}
( \ \rho_{n,x(t)} \ {\cal U}(t) \ \rho_{m,x(0)} \ ) \\
\label{Pnm}
P(n|m) \ \  & = & \ \mbox{trace}
( \ \rho_{n,x(t)} \ \rho_{m,x(0)} \ )
\end{eqnarray}
The phase-space propagator is denoted by ${\cal U}(t)$. 
The parametric kernel $P(n|m)$ depends on the displacement 
$\delta x$ but not on the actual time that it takes 
to realize this displacement. The trace operation is 
just a $dQdP$ integral over phase-space.

Before we go to the quantum mechanical analysis, 
let us summarize the classical scenario. 
The classical sudden approximation is 
\begin{eqnarray} 
P_t(n|m) \ \approx \ P(n|m)
\ \ \ \ \ \ \mbox{for $t \ll \tau_{\tbox{cl}}$}
\end{eqnarray}
For longer times we have the classical adiabatic 
approximation, or more precisely we have 
diffusive spreading:  
\begin{eqnarray}
P_t(n|m) \ \approx \ \mbox{Gaussian}(n{-}m)
\ \ \ \ \ \ \mbox{for $\tau_{\tbox{cl}} \ll t \ll t_{\tbox{frc}}$}
\end{eqnarray}
For $t \gg t_{\tbox{frc}}$ the kernel $P_t(n|m)$ is no-longer 
a narrow Gaussian that is centered around $n=m$. 
However, we can argue \cite{jar,frc} that its profile can be 
obtained as the solution of a stochastic 
diffusion equation (see Sec.9).

The kernel $P(n|m)$ reflects the parametric correlations 
between two sets of energy surfaces. 
Consequently Non-Gaussian features may manifest themselves.  
An important special non-Gaussian feature is encountered 
in many specific examples where $x$ affects only a tiny portion 
of the energy surface (Fig.\ref{f_surfaces}).
In the `piston' example this is the case 
because $({\partial {\cal H}}/{\partial x}) = 0$ unless 
$Q$ is near the face of the piston. Consequently 
$P(n|m)$ has a $\delta$-singularity for $n=m$. 

\section{Limitations on quantal-classical correspondence (QCC)} 

The main objects of our discussion are the 
transition probability kernel $P_t(n|m)$ and 
the parametric kernel $P(n|m)$ which have been 
introduced in the previous section. We refer 
now to Equations (\ref{Ptnm}) and (\ref{Pnm}).  
In the classical context $\rho_{n,x}(Q,P)$ is defined as 
the microcanonical distribution that is supported 
by the energy-surface ${\cal H}(Q,P;x(t))=E_n$. 
The energy $E_n$ corresponds to the phase-space 
volume $n{=}\Omega(E)$. In the QM context 
$\rho_{n,x}(Q,P)$ is defined as the Wigner-function 
that represents the energy-eigenstate $|n(x)\rangle$. 
The phase-space propagator is denoted 
symbolically by ${\cal U}(t)$. In the classical case 
it simply re-positions points in phase-space. 
In the QM case it has a more complicated structure.  
It is convenient to measure phase-space volume 
($n{=}\Omega(E)$) in units of $(2\pi\hbar)^d$ 
where $d$ is the number of degrees of freedom.  
This way we can obtain a `classical approximation' for 
the QM kernel, simply by making $n$ and $m$ integer 
variables. If the `classical approximation' is  
similar to the QM kernel, then we say that there is 
{\em detailed} QCC. If only the second-moment is 
similar, then we say that there is {\em restricted} QCC. 
In the present section we are going to discuss the 
conditions for having detailed QCC, using simple  
{\em semiclassical} considerations.

Wigner function $\rho_{n,x}(Q,P)$, unlike its classical 
microcanonical analog, has a non-trivial 
transverse structure. For a curved energy surface 
the transverse profile looks like Airy function and 
it is characterized by a width \cite{airy}
\begin{eqnarray} 
\Delta_{\tbox{SC}} \ = \ 
\left( \varepsilon_{\tbox{cl}} 
\left( \frac{\hbar}{\tau_{\tbox{cl}}} 
\right)^2 \right)^{1/3} 
\end{eqnarray}
where $\varepsilon_{\tbox{cl}}$ is a classical 
energy scale. For the `piston' example  
$\varepsilon_{\tbox{cl}}=E$ is the kinetic 
energy of the gas particle. In the next paragraph 
we discuss the conditions for having detailed QCC 
in the computation of the parametric kernel $P(n|m)$. 
Then we discuss the further restrictions on 
QCC, that are associated with the actual kernel $P_t(n|m)$.

Given a parametric change $\delta x$ we can define 
a classical energy scale 
$\delta E_{\tbox{cl}} \propto \delta x$ 
via (\ref{e25}).  This parametric 
energy scale characterizes the transverse 
distance between the intersecting energy-surfaces 
$|m(x)\rangle$ and $|n(x{+}\delta x)\rangle$. 
In the generic case, it should be legitimate to 
neglect the transverse profile of Wigner function 
provided $\delta E_{\tbox{cl}} \gg \Delta_{\tbox{SC}}$. 
This condition can be cast into the form 
$\delta x \gg \delta x_{\tbox{SC}}$ where 
\begin{eqnarray} 
\delta x_{\tbox{SC}} \  =  \  
\frac{\Delta_{\tbox{SC}}}
{\sqrt{\nu_{\tbox{E}}^{\tbox{cl}}/\tau_{\tbox{cl}}}}
\ \propto \ \hbar^{2/3}
\end{eqnarray}
Another important parametric scale 
is defined in a similar fashion: 
We shall see that it is {\em not} 
legitimate to ignore the transverse profile 
of Wigner function if 
$\delta E_{\tbox{cl}} < \Delta_b$.
This latter condition can be cast into the form 
$\delta x \ll \delta x_{\tbox{prt}}$ where  
\begin{eqnarray} 
\delta x_{\tbox{prt}} \  =  \  
\frac{\Delta_b}
{\sqrt{\nu_{\tbox{E}}^{\tbox{cl}}/\tau_{\tbox{cl}}}}
\ = \ 
\frac{\hbar}
{\sqrt{\nu_{\tbox{E}}^{\tbox{cl}} \tau_{\tbox{cl}}}} 
\end{eqnarray}
Typically the two parametric scales are well separated 
($\delta x_{\tbox{prt}} \ll \delta x_{\tbox{SC}}$). 
If we have $\delta x \ll \delta x_{\tbox{prt}}$ then the 
parametric kernel $P(n|m)$ is characterized by a 
perturbative core-tail structure 
which is illustrated in Fig.\ref{f_surfaces} and 
further discussed in the next sections. 
We do not have a theory for the 
intermediate parametric regime 
$\delta x_{\tbox{prt}} \ll \delta x \ll \delta x_{\tbox{prt}}$. 
But for $\delta x \gg \delta x_{\tbox{SC}}$ we 
can argue that there is a detailed 
QCC between the quantal kernel 
and the classical kernel.  Obviously, 
`detailed QCC' does not mean complete similarity.  
The classical kernel is typically characterized by 
various non-Gaussian features, such as sharp cutoffs, 
delta-singularities and cusps. These features are expected 
to be smeared in the quantum-mechanical case.

We turn now to discuss the actual transition 
probability kernel $P_t(n|m)$. Here we encounter 
a new restriction on detailed QCC:  
The evolving surface ${\cal U}(t)|m\rangle$ becomes more 
and more convoluted as a function of time. 
This is because of the mixing behavior that  
characterizes chaotic dynamics. For $t>t_{\tbox{scl}}$ 
the intersections with a given instantaneous energy 
surface $|n\rangle$ become very dense, and associated 
quantum-mechanical features can no longer be ignored. 
The time scale $t_{\tbox{scl}}$ can be related to the failure 
of the stationary phase approximation \cite{heller}.

The breaktime scale $t_{\tbox{scl}}$ of the 
semiclassical theory is analogous to 
the breaktime scale $t_{\tbox{prt}}$ of perturbation 
theory, as well as to the breaktime scale 
$t_{\tbox{frc}}$ of the classical theory.
In order to establish the crossover from 
ballistic to diffusive energy spreading 
using QCC considerations we should satisfy 
the condition $\tau _{\tbox{cl}} \ll t_{\tbox{scl}}$. 
This velocity-independent condition is not very restrictive. 
On the other hand we should also satisfy 
the condition $\delta x\gg \delta x_{\tbox{SC}}$, 
with $\delta x = V\tau _{\tbox{cl}}$. 
The latter condition implies that the applicability 
of the QCC considerations is restricted to relatively 
fast velocities. We can define: 
\begin{eqnarray}
v_{\tbox{SC}} \  =  \ 
\mboxs{scaled velocity} \ = \ 
\sqrt{ D_{\tbox{E}} \ \tau_{\tbox{cl}} } 
\ / \ \Delta_{\tbox{SC}}
\end{eqnarray}
If $v_{\tbox{SC}}\gg 1$ then the classical approximation
is applicable in order to analyze the crossover from 
ballistic to diffusive energy spreading.

\section{The parametric evolution of $P(n|m)$}

Detailed QCC between the quantal $P(n|m)$ 
and the classical $P(n|m)$ is not guaranteed if 
$\delta x < \delta x_{\tbox{SC}}$. For sufficiently small 
parametric change $\delta x$, perturbation theory becomes a 
useful tool for the analysis of this kernel. 
A detailed formulation of perturbation theory is postponed 
to later sections. Here we are going to sketch the main 
observations. We are going to argue that for small 
$\delta x$ there is no {\em detailed} QCC between the 
quantal and the classical kernels, but there is still 
{\em restricted} QCC that pertains to the second moment 
of the distribution.  Only for large enough  $\delta x$ 
we get detailed QCC. These observations are easily 
extended to the case of $P_t(n|m)$ in the next section.

For extremely small $\delta x$  the parametric kernel $P(n|m)$ 
has a standard ``first-order'' perturbative structure, namely: 
\begin{eqnarray}
P(n|m) \ \approx \ \delta_{nm} 
\ + \ \mbox{Tail}(n{-}m)
\ \ \ \ \ \mbox{for $\delta x \ll \delta x_c^{\tbox{qm}}$} 
\end{eqnarray}
where $\delta x_c^{\tbox{qm}}$ is defined as parametric 
change that is needed in order to mix neighboring levels. 
For larger values of $\delta x$ neighbor levels 
are mixed non-perturbatively and consequently we have 
a more complicated spreading profile: 
\begin{eqnarray}
P(n|m) \ \approx \ \mbox{Core}(n{-}m)
\ + \ \mbox{Tail}(n{-}m)
\ \ \ \ \ \mbox{for $\delta x \ll \delta x_{\tbox{prt}}$} 
\end{eqnarray}
In the perturbative regime ($\delta x \ll \delta x_{\tbox{prt}}$) 
the second moment of $P(n|m)$ is generically dominated by the `tail'. 
It turns out that the quantum-mechanical expression for the 
second-moment is classical look-alike, and consequently 
{\em restricted} QCC is satisfied. 
The {\em core} of the quantal $P(n|m)$ is of non-perturbative 
nature. The {\em core} is the component that is expected to become 
similar (eventually) to the classical $P(n|m)$.  
A large perturbation $\delta x \gg \delta x_{\tbox{prt}}$ 
makes the {\em core} spill over the perturbative {\em tail}.  
If we have also $\delta x \gg \delta x_{\tbox{SC}}$, then we 
can rely on detailed QCC in order to estimate $P(n|m)$.

For the piston example we can easily get estimates for the 
various scales involved. These are expressible in terms of 
De-Broglie wavelength 
$\lambda_{\tbox{E}}=2\pi\hbar/(\mathsf{m}v_{\tbox{E}})$ 
where $\mathsf{m}$ and  $v_{\tbox{E}}$ are defined as in Section 10.    
The displacement which is needed in order to mix neighboring levels, 
and the displacement which is needed in order to mix 
core and tail, are respectively:  
\begin{eqnarray} 
\delta x_c^{\tbox{qm}} \ \ \approx \ \ 
(\lambda_{\tbox{E}}^{d{+}1}/\mbox{\small Area})^{\tbox{1/2}}
\ \ \ll \ \ \lambda_{\tbox{E}}
\\
\delta x_{\tbox{prt}} \ \ \approx \ \ 
(\tau_{\tbox{col}}/{\tau_{\tbox{cl}}})^{\tbox{1/2}}
\ \lambda_{\tbox{E}} 
\ \ \gg \ \ \lambda_{\tbox{E}}
\end{eqnarray}
We have, in the generic case as well as in the case 
of the `piston', the hierarchy 
$\delta x_c^{\tbox{qm}} \ll \delta x_{\tbox{prt}}
\ll \delta x_{\tbox{SC}}$. Thus there is a `gap' between 
the perturbative regime ($\delta x \ll \delta x_{\tbox{prt}}$)
and the semiclassical regime ($\delta x \gg \delta x_{\tbox{SC}}$).

\section{The time evolution of $P_t(n|m)$}

\begin{figure}
\begin{center}
\leavevmode 
\epsfysize=3.0in 
\epsffile{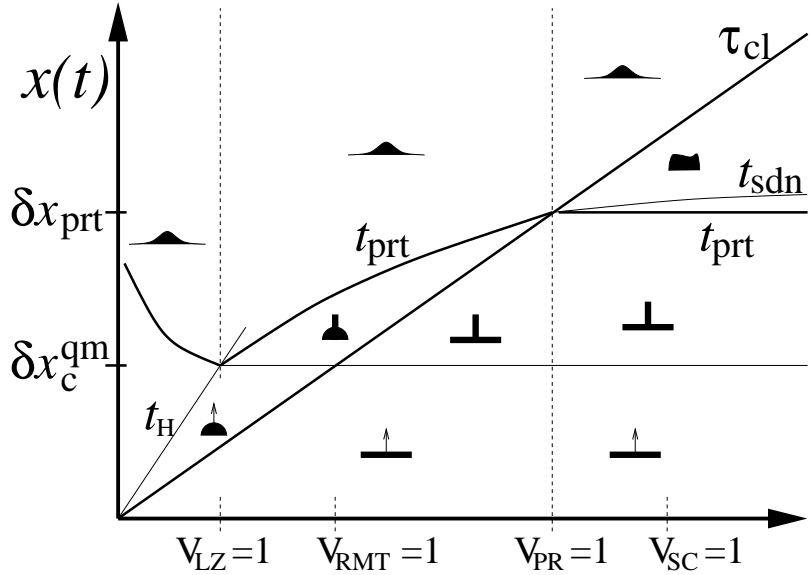}
\end{center}
\caption{\protect\footnotesize 
The various crossovers in the time evolution 
of $P_t(n|m)$. The vertical axis is $x(t)=Vt$.
The parametric scales $\delta x_c^{\tbox{qm}}$ 
and $\delta x_{\tbox{prt}}$ are indicted by 
horizontal lines. The horizontal  
axis is the velocity $V$. 
It is divided by vertical dashed lines 
to various velocity regimes. In each 
velocity regime there is a different dynamical 
route. The various crossovers are explained 
in the text and the various symbols are easily 
associated with having either Gaussian or 
some non-Gaussian spreading profile. In particular 
the perturbative spreading profile is either 
with or without non-trivial core, and its tail 
is either band-limited or resonance-limited.   
} 
\label{f_regimes}
\end{figure}

The dynamical evolution of $P_t(n|m)$ is related 
to the associated parametric evolution of $P(n|m)$.  
We can define a perturbative time scale $t_{\tbox{prt}}$ 
which is analogous to $\delta x_{\tbox{prt}}$.  
For $t\ll t_{\tbox{prt}}$ the kernel $P_t(n|m)$ is characterized 
by a core-tail structure that can be analyzed using perturbation theory.
In particular we can determine the second moment of 
the energy distribution, and we can establish {\em restricted} QCC. 
If the second moment for the core-tail structure is proportional 
to $t^2$, we shall say that there is a ballistic-like behavior. 
If it is proportional to $t$, we shall say that there is a 
diffusive-like behavior. {\em In both cases the actual 
energy distribution is not classical-like}, and therefore 
the term 'ballistic' and `diffusive' should be used with care. 
We are going now to give a brief overview of the various 
scenarios in the time evolution of $P_t(n|m)$. These are 
illustrated in Fig.\ref{f_regimes}. 
In later sections we give a detailed account of the theory.  

For {\em slow velocities} such that 
$\tau_{\tbox{cl}}\ll t_{\tbox{prt}} \ll t_{\tbox{H}}$, 
there is a crossover from ballistic-like spreading 
to diffusive-like spreading at $t\sim \tau_{\tbox{cl}}$. 
In spite of the lack of detailed QCC there is still 
restricted QCC as far as this ballistic-diffusive 
crossover is concerned. 
The breakdown of perturbation theory before 
the Heisenberg time ($t_{\tbox{prt}} \ll t_{\tbox{H}}$)  
implies that there is a second crossover 
at $t \sim t_{\tbox{prt}}$ from a diffusive-like spreading 
to a genuine diffusive behavior. 

{\em extremely slow velocities} are defined by the 
the inequality $t_{\tbox{H}} \ll t_{\tbox{prt}}$. This 
inequality implies that there are quantum-mechanical 
recurrences {\em before} the expected crossover from 
diffusive-like spreading  to genuine-diffusion. 
This is the quantum-mechanical adiabatic regime. In the 
$t\rightarrow\infty$ limit Landau-Zener transitions 
dominate the energy spreading, and consequently 
neither detailed nor restricted QCC is a-priori expected \cite{wilk}.   
 
For {\em fast velocities} we have  
$t_{\tbox{prt}} \ll \tau_{\tbox{cl}}$. 
There is a crossover at $t\sim t_{\tbox{prt}}$ from 
ballistic-like spreading to a genuine ballistic behavior, 
and at $t \sim \tau_{\tbox{cl}}$ there is a second  
crossover from genuine-ballistic to 
genuine-diffusive spreading. The description of  
this classical-type crossover is out-of-reach for 
perturbation theory, but we can use the semiclassical 
picture instead.  Note that 
the semiclassical definition of `fastness' and the 
perturbative definition of `slowness' imply 
that there is a `gap' between the corresponding regimes. 
However, the interpolation is smooth, and therefore 
for simple systems surprises are not expected.

\section{Linear response theory}

The classical derivation in Section 8 applies also 
in the quantum-mechanical case provided ${\cal F}(t)$ 
is treated as an operator. This is known as 
`linear response theory', and the expression for 
$D_{\tbox{E}}$ is essentially the `Kubo-Greenwood' formula. 
Obviously the restriction $t\ll t_{\tbox{frc}}$ is replaced by 
the more restrictive condition $t\ll t_{\tbox{prt}}$, 
which will be discussed later.
For the purpose of concise presentation, the formula 
for the energy spreading can be written as follows:
\begin{eqnarray} \label{e43} 
\delta E^2 \ = \  
V^2\int_0^t\int_0^t C_{\tbox{E}}(t_2{-}t_1)dt_1dt_2
\ = \ 
V^2 t \int_{-\infty}^{+\infty}\frac{d\omega}{2\pi}
\tilde{C}_{\tbox{E}}(\omega) \ \tilde{F}_t(\omega)
\end{eqnarray} 
\begin{eqnarray} \label{Fomega}
\mbox{where} \hspace{2cm}
\tilde{F}_t(\omega) \ = \ t{\cdot}(\mbox{sinc}(\omega t /2))^2
\end{eqnarray} 
The power spectrum of the classical fluctuations looks like 
white noise. It satisfies 
$\tilde{C}_{\tbox{E}}(\omega) \approx \nu_{\tbox{E}}$ 
for $|\omega|<1/\tau_{\tbox{cl}}$ and decays to zero 
outside of this regime. Thus, still considering the classical 
case, for $t \ll \tau_{\tbox{cl}}$ 
we can make the replacement $\tilde{F}_t(\omega)\rightarrow t$, 
and we obtain the ballistic result 
$\delta E^2 = C_{\tbox{E}}(0){\cdot}(Vt)^2$,  
while for  $t \gg \tau_{cl}$ we can make the replacement 
$\tilde{F}_t(\omega)\rightarrow 2\pi\delta(\omega)$, and 
we get then the diffusive behavior 
$\delta E^2 = \nu_{\tbox{E}}V^2 t$.
Now we turn to the quantum-mechanical case. 
The power-spectrum of the quantum-mechanical fluctuations 
is given by the formula
\begin{eqnarray} \label{Comega}
\tilde{C}_{\tbox{E}}(\omega) \ = \ 
\sum_n'
\left|\left(\frac{\partial {\cal H}}{\partial x}\right)_{nm}\right|^2
\ 2\pi\delta\left(\omega-\frac{E_n{-}E_m}{\hbar}\right)
\end{eqnarray} 
Semiclassical reasoning \cite{mario} applied to (\ref{Comega}) 
leads to the immediate conclusion that energy levels are coupled 
by matrix elements provided $|E_n-E_m|<\Delta_b$ where
\begin{eqnarray} 
\Delta_b \ = \ 
\frac{\hbar}{\tau_{\tbox{cl}}} \ = \ 
\mbox{band width}
\end{eqnarray} 
The discrete nature of the power spectrum is of no 
significance as long as  $t \ll t_{\tbox{H}}$. 
Therefore we have a crossover from ballistic 
to diffusive behavior as in the classical case.  
On the other hand, if $t \gg t_{\tbox{H}}$   
we have $\delta E^2 \sim \mbox{const}$. This is due 
to {\em quantum-mechanical recurrences} \cite{berry}.    
We shall argue that the latter result is valid only 
for {\em extremely slow velocities}, for which 
$t_{\tbox{H}} \ll t_{\tbox{prt}}$, and provided Landau-Zener transitions
are ignored.  This is the {\em quantum-mechanical adiabatic regime}.
For {\em non} extremely slow velocities we have  
$t_{\tbox{H}} \gg t_{\tbox{prt}}$, 
and consequently there is a second crossover 
at $t=t_{\tbox{prt}}$ from diffusive-like behavior to 
genuine diffusion. In the latter case 
there are no recurrences, and QCC holds also for~$t>t_{\tbox{H}}$.

\section{Actual and Parametric Dynamics}

The simplicity of linear response theory is lost once 
we try to formulate a controlled version of it. 
Therefore it is better to use a more conventional 
approach and to view the energy spreading as arising 
from {\em transitions between energy levels}. 
The transition probability kernel and the parametric kernel
can be written using standard Dirac notations as follows: 
\begin{eqnarray} 
P_t(n|m) \ = \ & |{\mathbf U}_{nm}(t)|^2 \ = \ &
\left|\langle n(x(t))| {\mathbf U}(t) | m(0) \rangle \right|^2
\\
P(n|m) \ = \ & |{\mathbf T}_{nm}(x)|^2 \ = \ & 
\left|\langle n(x)| m(0) \rangle \right|^2
\end{eqnarray} 
The evolution matrix ${\mathbf U}_{nm}(t)$ can be obtained 
by solving the Schroedinger equation 
\begin{eqnarray} \label{e_LD}
\frac{da_n}{dt} \ = \ 
-\frac{i}{\hbar}E_n \ a_n \ -  
\frac{i}{\hbar}\sum_m {\mathbf W}_{nm}(x(t)) \ a_m 
\\
\mbox{where} \hspace{2cm}
{\mathbf W}_{nm} \ = \ 
i\frac{\hbar\dot{x}}{E_n{-}E_m} 
\Big\langle n \Big| \frac{\partial{\cal H}}{\partial x} \Big| m \Big\rangle
\end{eqnarray}
The derivation of (\ref{e_LD}) follows standard procedure \cite{frc}.  
The transformation matrix ${\mathbf T}_{nm}$ can be 
obtained by considering the {\em same} equation with the 
first term on the right hand side omitted. 
We shall refer to ${\mathbf T}_{nm}(x(t))$ as describing 
the parametric dynamics (PD), while ${\mathbf U}_{nm}(t)$ 
describes the actual dynamics (AD).
For PD the velocity $\dot{x}=V$ plays no role, 
and it can be scaled out from the above 
equation. Consequently, for PD, parametric scales and 
temporal scales are trivially related via the scaling 
transformation  $\delta x = V \tau$. 
For short times, as long as the energy differences 
between the participating levels are not yet resolved,  
the AD coincided with the PD. This is the quantum-mechanical 
sudden approximation, which we are going to further discuss 
later on.

\begin{table}
\begin{center}
\leavevmode  
\large
\fbox{
\hspace*{1cm} 
\mpg{8}{\setlength{\baselineskip}{3cm}
\begin{tabular}{lllll}
\ & \ & \ & \ & \ \\ 
\multicolumn{5}{l}{\bf Classical Parameters:} \\
$\tau_{\tbox{cl}}$  & = & \mbox{\small correlation time} & \ & \ \\
$D_{\tbox{E}}$      & = & $\frac{1}{2} \ \nu_{\tbox{E}} \ V^2$ & \ & \ \\
\ & \ & \ & \ & \ \\ 
\multicolumn{5}{l}{\bf Quantum-Mechanical Parameters:} \\
$\Delta_b$  & = & \mbox{\small band width}    & \ & \ \\ 
$\Delta$    & = & \mbox{\small level spacing} & \ & \ \\ 
\ & \ & \ & \ & \ \\
\multicolumn{5}{l}{\bf Primary Dimensionless Parameters:} \\
$v_{\tbox{PR}}$  & = & \mbox{\small scaled velocity} & 
= & $\sqrt{D_{\tbox{E}} \ \tau_{\tbox{cl}}} \ / \ \Delta_b$ \\
$b$ & = & \mbox{\small scaled band width} & 
= & ${\Delta_b} \ / \ {\Delta}$ \\
\ & \ & \ & \ & \ \\
\multicolumn{5}{l}{\bf Secondary Dimensionless Parameters:} \\ 
$v_{\tbox{RMT}}$ & = & $b^{\tbox{1/2}} \ v_{\tbox{PR}}$ & 
= & $\tau_{\tbox{cl}} \ / \ \tau_c^{\tbox{qm}}$  \\
$v_{\tbox{LZ}}$ & =  & $b^{\tbox{3/2}} \ v_{\tbox{PR}}$ & 
= & $t_{\tbox{H}} \ / \ \tau_c^{\tbox{qm}}$  \\
\ & \ & \ & \ & \ 
\end{tabular}
} \hspace*{1cm} } 
\end{center}
\caption{\protect\rm\footnotesize 
Generic parameters in the classical and in the quantum-mechanical 
theories of dissipation. Note that quantum-mechanics requires the 
specification of two additional parameters. However, for the 
purpose of QCC considerations only $\Delta_b$ is significant. 
Note that $v_{\tbox{LZ}} \gg v_{\tbox{RMT}} \gg v_{\tbox{PR}}$.
In the classical limit all of them $\gg 1$. }
\end{table}

The generic parameters that appear in the quantum-mechanical 
theory are summarized in Table I.  The specification of 
the mean level spacing $\Delta$ is not dynamically significant 
as long as $t\ll t_{\tbox{H}}$.  Longer times are required in order 
to resolve individual energy levels. Thus we come to the 
conclusion that in the time regime $t<t_{\tbox{H}}$
there is a {\em single} generic dimensionless 
parameter, namely $v_{\tbox{PR}}$, 
that controls QCC.
We shall see that the quantum mechanical definition of slowness, 
namely $\tau_{\tbox{cl}} \ll t_{\tbox{prt}}$, can be cast into 
the form $v_{\tbox{PR}} \ll 1$.  On the other hand 
in the classical limit we have  
$v_{\tbox{PR}} \gg v_{\tbox{SC}} \gg 1$. 
Thus the dimensionless parameter $v_{\tbox{PR}}$ marks a border 
between two regimes where different considerations 
are required in order to establish QCC.

\section{Perturbation theory}

We can use Equation (\ref{e_LD}) as a starting point for 
a conventional first-order perturbation theory. 
For short times, such that $P_t(m|m) \sim 1$, 
the transition probability from level $m$ to level $n$ 
is determined by the coupling strength $|\mbf{W}_{nm}|^2$, 
by the energy difference $(E_n{-}E_m)$ and 
by the correlation function $F(\tau)$. 
The latter describes loss of correlation between 
$\mbf{W}_{nm}(x(0))$ and $\mbf{W}_{nm}(x(t))$. 
It is defined via 
\begin{eqnarray} \label{xcorr}
\left\langle
\ \mbf{W}_{nm}^{\star}(t{+}\tau)  
\ \mbf{W}_{nm}(t) \ 
\right\rangle \ = \ 
\ |W_{nm}|^2 \ F(\tau) 
\end{eqnarray} 
with the convention $F(0)=1$. It is now quite straightforward \cite{frc}
to obtain, using first-order perturbation theory, the following result:  
\begin{eqnarray} \label{e_perth}
P_t(n|m) \ \approx \ 
t\tilde{F}_t\left(\frac{E_n{-}E_m}{\hbar}\right) \times 
\left| \frac{\mbf{W}_{nm}}{\hbar} \right|^2  
\ \ \ \ \ \ \mbox{for $n\ne m$} \ \ \ .
\end{eqnarray} 
The function $\tilde{F}_t(\omega)$ describes the 
spectral content of the perturbation. For a {\em constant}  
perturbation ($F(\tau)=1$) it is just given 
by equation (\ref{Fomega}). For a {\em noisy} perturbation 
$F(\tau)$ is characterized by some finite correlation-time 
$\tau_c$, and therefore  the function  $\tilde{F}_t(\omega)$
is modified as follows:
\begin{eqnarray} 
\tilde{F}_t(\omega) \ = \ \tilde{F}(\omega) 
\ \ \ \ \mbox{for $t>\tau_c$}
\end{eqnarray} 
where $\tilde{F}(\omega)$ is the Fourier transform of 
the correlation function $F(\tau)$. 
In order to use (\ref{e_perth}) we should determine how 
$F(\tau)$ look like, and in particular we should determine 
what is the correlation-time $\tau_c$. We postpone 
this discussion, and assume that $F(\tau)$ and hence 
$\tau_c$ are known from some calculation. The total 
transition probability is $p(t)=\sum_n'P(n|m)$, where
the prime indicates omission of the term $n=m$. 
First order perturbation theory is valid as long as 
$p(t)\ll 1$. This defines the breaktime $t_{\tbox{prt}}'$  
of the perturbative treatment. We use the notation 
$t_{\tbox{prt}}'$ rather than $t_{\tbox{prt}}$, since later we are 
going to define an improved perturbative treatment 
where $t_{\tbox{prt}}$ is defined differently.

Using the above first-order perturbative result  
we can obtain the previously discussed expression (\ref{e43}) 
for the energy spreading:
\begin{eqnarray} \label{e_43a}
\delta E^2 \ = \  
\sum_n (E_n{-}E_m)^2 \ P_t(n|m) \ = \ 
V^2 t \int_{-\infty}^{+\infty}\frac{d\omega}{2\pi}
\tilde{C}_{\tbox{E}}(\omega) \ \tilde{F}_t(\omega)
\end{eqnarray} 
We see that the perturbative result would coincides 
with the linear-response result only if the coupling 
matrix-elements could have been treated as {\em constant} in time.
The above derivation imply that we can trust this 
formula only for a short time $t<t_{\tbox{prt}}'$. 
However, later we shall see that it can be 
trusted for a longer time $t<t_{\tbox{prt}}$.

It is now possible to formulate the conditions 
for having restricted QCC. 
By `restricted' QCC we mean that only $\delta E^2$ is 
being considered. We assume that the 
result (\ref{e_43a}) is valid for $t<t_{\tbox{prt}}$. 
We also assume that $t_{\tbox{prt}}$ as well as 
$F(\tau)$ and hence $\tau_c$ are known from some calculation.  
The following discussion 
is meaningful if and only if the following 
Fermi-golden-rule (FGR) condition is satisfied:  
\begin{eqnarray} \label{FGRc}
\mbox{\bf FGR-condition:} \ \ \ \  
\mbox{Either} \ \tau_{\tbox{cl}} \ \mbox{or} \ \tau_c  
\ \ \ \ll \ \ \ t_{\tbox{prt}}   
\end{eqnarray} 
It is essential to distinguish between 
two different possible scenarios:
\begin{eqnarray}
\mbox{Resonance-limited transitions:} \ \ \ \ \ &
\tau_c \gg \tau_{\tbox{cl}} \\
\mbox{Band-limited transitions:} \ \ \ \ \ &
\tau_c \ll \tau_{\tbox{cl}} 
\end{eqnarray}
For resonance-limited transitions,  
finite $\tau_c$ has no consequence as far as 
$\delta E^2$ is concerned: The crossover 
to diffusive behavior $\delta E^2 \propto t$ 
happens at $t\sim\tau_{\tbox{cl}}$, and this 
diffusive behavior persists for $t>\tau_c$ with 
the same diffusion coefficient. On the other hand, for 
band-limited transitions we have at 
$t\sim\tau_c$  a pre-mature crossover from 
ballistic to diffusive behavior. 
Consequently the classical result is  
suppressed by a factor $(\tau_c/\tau_{\tbox{cl}}) \ll 1$. 
This is due to the fact that the transitions 
between levels are limited not by 
the resonance width (embodied~by~$\tilde{F}(\omega)$), 
but rather by the band-width of the coupling matrix elements 
(embodied~by~$\tilde{C}_{\tbox{E}}(\omega)$).

\section{The over-simplified RMT picture}

In order to have practical estimates for 
$t_{\tbox{prt}}'$ and $\tau_c$ it is required 
to study the statistical properties of 
the matrix ${\mathbf W}_{nm}$.  This matrix 
is banded, and its elements satisfy:
\begin{eqnarray}
\left\langle 
\left|\frac{{\mathbf W}_{nm}}{\hbar}\right|^2
\right\rangle 
\ \approx \ 
\left(\frac{1}{\tau_c^{\tbox{qm}}}\right)^2 
\ \frac{1}{(n{-}m)^2} 
\ \ \ \ \ \mbox{for} \ \ |n{-}m|<b/2 
\end{eqnarray}
The time scale $\tau_c^{\tbox{qm}}$ is related 
via $\delta x = V \tau$ to the parametric scale 
\begin{eqnarray}
\delta x_c^{\tbox{qm}} \ = \ 
\frac{\Delta}{\sqrt{\langle |
({\partial {\cal H}}/{\partial x})_{nm}
|^2\rangle }} \ \approx \ 
\sqrt{\frac{2\pi\hbar\Delta}{\nu_{\tbox{E}}^{\tbox{cl}}}} 
\end{eqnarray}
This is the parametric change which is required 
in order to mix neighboring levels. 
Given two levels $n$ and $m$ one observes that 
$\delta x_c^{\tbox{qm}}$ also determines  
the correlation scale of the matrix-element 
${\mathbf W}_{nm}(x(t))$. 
These observations can be summarized as follows: 
\begin{eqnarray} \label{e_simple}
t_{\tbox{prt}}' \ = \ \tau_c \ = \ \tau_c^{\tbox{qm}} 
\ \ \ \ \ \mbox{for standard perturbation theory.}  
\end{eqnarray}
In the regime  $v_{\tbox{RMT}}\ll 1$ we have  
$\tau_c  \gg  \tau_{\tbox{cl}}$. 
Therefore we have resonance limited transitions 
and we can relay on (\ref{e_43a}) in order 
to establish restricted QCC.    
On the other hand in the regime $v_{\tbox{RMT}}\gg 1$ 
the FGR condition (\ref{FGRc}) is not satisfied,  
and the expected crossover at $t=\tau_{\tbox{cl}}$ is 
out-of-reach as far as the standard version of 
perturbation theory is concerned.

In the spirit of random-matrix-theory (RMT) we may 
think of ${\mathbf W}_{nm}(x)$ of Equation (\ref{e_LD}) 
as a particular realization which is taken out from 
some large ensemble of (banded) random matrices. 
In order to have a well defined mathematical model 
we should specify the $x$-correlations as well. 
It looks quite innocent to assume that the only 
significant correlations are those expressed in 
Eq.(\ref{xcorr}). In other words, let us assume, 
following \cite{WA}, that 
{\em cross correlations between matrix elements} 
can be ignored.  One finds then that (\ref{e_43a}) 
should hold for classically  long times.  
This claim can be summarized as follows: 
\begin{eqnarray} \label{e_RMT}
t_{\tbox{prt}} = t_{\tbox{frc}}
\ \ \ \ \ \ \mbox{and} \ \ \ \ \ \ 
\tau_c = \tau_c^{\tbox{qm}} 
\ \ \ \ \ \ \mbox{for the over-simplified RMT picture.} 
\end{eqnarray}
If the over-simplified RMT assumption were true it would 
imply that in the $v_{\tbox{RMT}}\gg 1$ regime the 
transitions would be band-limited and consequently 
classical diffusion would be suppressed by a factor 
$(1/v_{\tbox{RMT}})\ll 1$.  Note that in the 
semiclassical limit we have indeed $v_{\tbox{RMT}}\gg 1$. 
Therefore, it is implied that the 
classical limit would not coincide with the 
classical result. Obviously, we expect this conclusion 
to be wrong, and indeed we shall demonstrate that 
{\em cross correlations between matrix elements 
cannot be ignored}. This will be done by transforming 
the Schroedinger equation to a more appropriate basis.

\section{The perturbative core-tail spreading profile}

\begin{figure}[h]
\begin{center}
\leavevmode  
\epsfysize=1.5in 
\epsffile{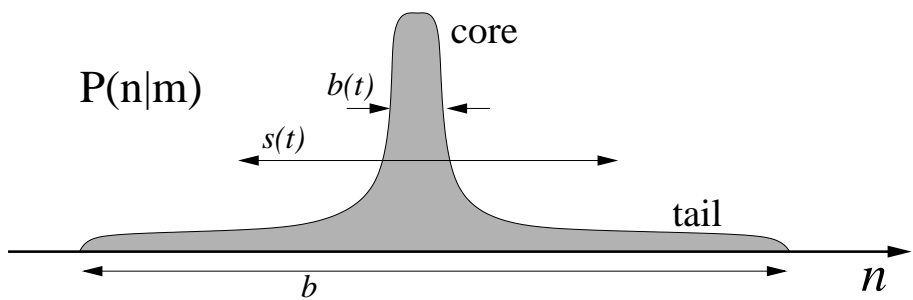}
\end{center}
\caption{\protect\footnotesize  
Schematic illustration of a generic core-tail spreading profile.
The core-width $b(t)$ is defined by the participation-ratio. 
The second-moment should satisfy $b(t) \ll s(t) \ll b$, 
where $b$ is that bandwidth. In case of $P_t(n|m)$ the 
tail becomes  (for $t>\tau_{\tbox{cl}}$) resonance-limited 
rather than band-limited.  In the resonance-limited case 
the bandwidth $b$ in the above figure should be replaced 
by $(\hbar/t)/\Delta$, and accordingly the requirement is 
$b(t) \ll s(t) \ll (\hbar/t)/\Delta$.}
\label{f_coretail}
\end{figure}

The standard version of perturbation theory is valid 
for an extremely short time. See (\ref{e_simple}).   
In order to formulate an improved version of  perturbation 
theory it is essential to understand the perturbative structure 
of $P_t(n|m)$. Once $t>\tau_c^{\tbox{qm}}$ neighboring levels are 
mixed and consequently the probability kernel acquires a 
non-trivial core-tail structure which is illustrated 
in Fig.\ref{f_coretail}. The `improved' version of perturbation 
theory should be applicable as long as the core-tail structure 
is maintained. 

Now we shall characterize the main features 
of a generic core-tail spreading profile. 
The expression for the probability kernel $P_t(n|m)$ can 
be written schematically as follows:
\begin{eqnarray}
P_t(n|m) \ \approx \ \mbox{Core}(n{-}m) 
\ + \ \mbox{Tail}(n{-}m) 
\ \ \ \ \ \mbox{for $t <  t_{\tbox{prt}}$}
\end{eqnarray}
The kernel is characterized by two scales:
\begin{eqnarray}
b(t) \ = \ \mbox{\small core width} \ &=& \ 
\left( \sum_n (P_t(n|m))^2 \right)^{-1}  \\
s(t) \ = \ \mbox{\small spreading} \ &=& \
\left( \sum_n (n{-}m)^2 \ P_t(n|m) \right)^{-1/2}
\end{eqnarray}
such that $b(t) \ll s(t) \ll b$. 
For $t\ll\tau_c^{\tbox{qm}}$ we have a trivial 
core with $b(t) \approx 1$, whereas for $t\gg\tau_c^{\tbox{qm}}$
we have a non-trivial core with $b(t)\gg 1$.  
The matrix elements satisfy 
$\langle |{\mathbf W}_{nm}|^2 \rangle 
\ \propto \ 1/(n{-}m)^2$. 
We shall see (see (\ref{e73})) that in the `band-limited tail' case 
we have $P_t(n|m) \sim \mbox{const}/(n{-}m)^2$  
up to the cutoff $b$, while for the `resonance-limited tail'
case we have $P_t(n|m) \sim \mbox{const}/(n{-}m)^2$ 
up to the cutoff $(\hbar/t)/\Delta$.  One should realize  
that the power-law behavior of the tail is `fast' enough
in order to guarantee that $b(t)$ is independent of the  
tail's cutoff.  The cutoff does not have any effect on 
the evolving core. On the other hand, the second moment $s(t)$, 
unlike $b(t)$, is predominantly determined by the tail's cutoff, 
and it is independent of the core structure.

\section{An improved perturbation theory}

In order to extend perturbation theory beyond 
$\tau_c^{\tbox{qm}}$ it is essential to eliminate 
the non-perturbative transitions within the core. 
This can be done by making a transformation to 
an appropriate basis as follows:
\begin{eqnarray}
a_n(t) \ &=& \ \sum_m \tilde{\mathbf{T}}_{nm} \ c_m(t) \\
\tilde{\mathbf{T}}_{nm}  &=&  {\mathbf{T}}_{nm}
\ \ \ \mbox{if $|n{-}m|<b'/2$, else zero.} 
\end{eqnarray}
The amplitudes $c_n(t)$ satisfy the same Schroedinger 
equation as the $a_n(t)$, with a transformed matrix
$\tilde{\mathbf{W}}$. The general expression 
for $\tilde{\mathbf{W}}$ is quite complicated, 
but we are interested only in the core-to-tail 
transitions for which 
\begin{eqnarray} 
(\tilde{\mathbf{W}})_{nm}=
(\tilde{\mathbf{T}}^{\dagger}\mathbf{W}\tilde{\mathbf{T}})_{nm}
\ \ \ \mbox{for $|n{-}m|>b'$}
\end{eqnarray}
(no approximation is involved). 
Once this transformation is performed the `new' Schroedinger 
equation is characterized by a new correlation 
time $\tau_c$ and by a new perturbative time $t_{\tbox{prt}}'$. 
Both $\tau_c$ and $t_{\tbox{prt}}'$ depend on the 
free parameter $b'$. Our choice of the 
course-graining parameter $b'$ is not completely arbitrary.  
The restrictions are: \\  
\ \\
$\hspace*{2mm} \bullet \ \ $ 
Unitarity is approximately preserved: 
\ \ \ \ \ \ \ \ \ $b(t) \ \ll \ b'$. \\
$\hspace*{2mm} \bullet \ \ $ 
Core-to-Tail transitions are preserved:
\ \ \ \ \ \ \ \ \ $b' \ \ll \ b$. \\
$\hspace*{2mm} \bullet \ \ $ 
Long effective correlation time is attained:
\ \ \ \ $\tau_{\tbox{cl}} \ \ll \ \tau_c$ \\
\ \\
The feasibility of the last requirement deserves further 
discussion. One should realize that for $b'=b$ we are actaully 
transforming the Schrodinger equation to an $x$-independent basis. 
Therefore the transformed matrix 
$\tilde{\mathbf{T}}^{\dagger}\mathbf{W}\tilde{\mathbf{T}}$
becomes correlated on a time scale $\tau_c^{\tbox{cl}}$, 
which has been discussed in Section 5. 
As we change $b'$ from $b$ to smaller values, 
we expect $\tau_c$ to become smaller. By continuity, we 
expect no difficulty in satisftying the conditions 
$b'\ll b$ and $\tau_c\gg\tau_{\tbox{cl}}$ simultaneously.

The usefullness of the above transformation stems from 
the fact that due to the elimination of non-perturbative 
transitions within the core, $t_{\tbox{prt}}'$  becomes much 
longer than $\tau_c^{\tbox{qm}}$. At the same time the 
information which is required in order to determine 
the second moment $s(t)$ is not lost. 
We have $|\tilde{\mathbf{W}}_{nm}| \approx |{\mathbf{W}}_{nm}|$
for core-to-tail transitions, and a practical approximation 
for the `renormalized' spreading profile would be 
\begin{eqnarray} \label{e73}
P_t(n|m) \ \sim \ \delta_{nm} \ + \ 
t \tilde{F}_t\left( \frac{E_n{-}E_m}{\hbar}\right)
\times\left(\frac{1}{\tau_c^{\tbox{qm}}}\right)^2 
\ \frac{1}{(b')^2+(n{-}m)^2}   
\end{eqnarray}
The behavior for $|n{-}m| \le b'$ is an artifact of 
the transformation and contains false information. 
However, for the calculation of the second moment 
only the tail is significant. The tail is not affected
by our transformation and therefore we 
obtain the {\em same} result (\ref{e_43a}) for $\delta E^2$ with 
one important modification: a different effective 
value for $\tau_c$. Moreover, since $b'$ is 
chosen such that $\tau_c\gg\tau_{\tbox{cl}}$, it follows 
that the transitions are resonant-limited and consequently 
restricted QCC is established also 
in the domain $v_{\tbox{RMT}}>1$. 

The validity of the above QCC considerations is conditioned 
by having $\tau_{\tbox{cl}} \ll t_{\tbox{prt}}'$.  
Breakdown of the present version of perturbation theory 
happens once the total transition probability 
in (\ref{e73}) becomes non-negligible (of order 1). Thus   
\begin{eqnarray} 
t_{\tbox{prt}}' \ = \ (b')^{\tbox{1/2}} \times \tau_c^{\tbox{qm}} 
\ \ \ \ \ \mbox{for the improved perturbation theory.}
\end{eqnarray}
It is easily verified that the latter condition 
$\tau_{\tbox{cl}} \ll t_{\tbox{prt}}'$ cannot be 
satisfied if $v_{\tbox{PR}}>1$.  This is not just a technical 
limitation of our perturbation theory, but reflects a real 
difference between two distinct routes towards QCC. 
This point is further illuminated in the next sections.

\section{Consequences of the improved perturbative treatment}

Our perturbation theory is capable of 
giving information about the tail, and 
hence about the second moment. 
Given $t$, one wonders how much $b'$  can 
be `pushed down' without violating 
the validity conditions of our procedure. 
It is quite clear that $b'\gg b(t)$ is 
a necessary condition for {\em not} having a breakdown   
of perturbation theory. If we {\em assume} that the 
energy-spreading-profile is characterized just by 
the single parameter $b(t)$, then the condition 
$b'\gg b(t)$ should be equivalent to $t\ll t_{\tbox{prt}}'$.  
Hence the following estimate is suggested:  
\begin{eqnarray}
b(t) \ = \ (t/\tau_c^{\tbox{qm}})^2
\end{eqnarray}
If we want to have a better idea about 
the core structure we should apply, in any 
special example, specific (non-perturbative) considerations. 
For example, in case of the piston example 
we can use semiclassical considerations in order 
to argue \cite{wls} that the core has a Lorentzian shape 
whose width is $\hbar/\tau_{\tbox{col}}$. 
This structure is exposed provided  
$\hbar/\tau_{\tbox{col}} \ll b(t)$,  
leading to the condition $\delta x \gg \lambda_{\tbox{E}}$. 
Else we have a structure-less core whose 
width is characterized by the single parameter $b(t)$.

We turn now to determine the $\delta x_{\tbox{prt}}$ of the 
parametric dynamics (PD), and then the $t_{\tbox{prt}}$ of 
the actual dynamics (AD).
Recall that PD is obtained formally by 
ignoring the differences $(E_n{-}E_m)$, 
which implies that we can 
make in (\ref{e73}) the replacement 
$\tilde{F}_t \mapsto t$.  Thus the 
tail of $P(n|m)$ is band-limited and 
consequently the second moment is 
\begin{eqnarray}
s(t)^2 \ = \ b \times (1/\tau_c^{\tbox{qm}})^2 \ t^2
\ \ \ \mbox{[band-limited tail]}
\end{eqnarray}
in agreement with the classical ballistic result. 
Our procedure for analyzing the core-tail structure 
of $P(n|m)$ is meaningful as long as 
we have $b(t) \ll s(t) \ll b$. This defines 
an upper time limitation 
$t_{\tbox{prt}}=\delta x_{\tbox{prt}}/V$, where 
\begin{eqnarray}
\delta x_{\tbox{prt}} \ = \
b^{\tbox{1/2}} \ \delta x_c^{\tbox{qm}} \ = \ 
\frac{\hbar}{\sqrt{\nu_{\tbox{E}}^{\tbox{cl}}
\tau_{\tbox{cl}}}}
\end{eqnarray}
At $t=t_{\tbox{prt}}$ we have $b(t) \sim s(t) \sim b$, 
and we expect a crossover from a ballistic-like 
spreading to a genuine  ballistic spreading.

The AD departs from the PD once the energy scale 
$\Delta_b$ is resolved. 
This happens when $t\sim\tau_{\tbox{cl}}$. 
The perturbative approach is applicable for the 
analysis of the crossover at $t\sim\tau_{\tbox{cl}}$ 
provided  $V\tau_{\tbox{cl}}\ll\delta x_{\tbox{prt}}$. 
This is precisely the condition $v_{\tbox{PR}}\ll 1$. 
For $t\gg\tau_{\tbox{cl}}$ the tail becomes resonance 
limited ($|n{-}m|<(\hbar/t)/\Delta$) 
rather than band limited  ($|n{-}m|<b$) and  we obtain:
\begin{eqnarray} 
s(t)^2 \ = \ 
(1/\tau_c^{\tbox{qm}})^2 \ t_{\tbox{H}} \ t 
\ \ \ \mbox{[resonance-limited tail]} 
\end{eqnarray}
in agreement with the classical diffusive result. 
Our procedure for analyzing the core-tail structure 
of $P_t(n|m)$ is meaningful as long as 
we have $b(t) \ll s(t) \ll b$. This defines 
a {\em modified} upper time limitation 
\begin{eqnarray}
t_{\tbox{prt}} \ = \ 
(\tau_c^{\tbox{qm}})^{\tbox{2/3}} \ t_{\tbox{H}}^{\tbox{1/3}} 
\ = \ 
\left( \frac{\hbar^2}
{\nu_{\tbox{E}}V^2}\right)^{1/3} 
\ \ \ \ \ \mbox{[applies to $v_{\tbox{PR}}\ll 1$]}
\end{eqnarray}
At $t=t_{\tbox{prt}}$ we have $b(t) \sim s(t) \ll b$, 
and we expect a crossover from a diffusive-like 
spreading to a genuine diffusive spreading.

\section{The quantum mechanical sudden approximation}

\begin{table}[h]
\begin{center}
\leavevmode  
\large
\fbox{
\hspace*{1cm} 
\mpg{9}{\setlength{\baselineskip}{3cm}
\begin{tabular}{ll}
\ & \ \\ 
\multicolumn{2}{l}{{\bf Perturbative route} ($v_{\tbox{PR}} \ll 1$):} \\
\ & \ \\
$t_{\tbox{sdn}} = \tau_{\tbox{cl}} \ll t_{\tbox{prt}}$ & \ \\ 
\ & \ \\
At \ \ $t = \tau_{\tbox{cl}}$ &
$b(t) \ll s(t) \ll b \sim (\hbar/t)/\Delta$  \\   
At \ \ $t = t_{\tbox{prt}}$  &  
$b(t) \sim s(t) \sim  (\hbar/t)/\Delta  \ll  b$ \\
\ & \ \\ 
\ & \ \\
\multicolumn{2}{l}{{\bf Non-perturbative route} ($v_{\tbox{PR}} \gg 1$):} \\ 
\ & \ \\
$t_{\tbox{prt}} \ll t_{\tbox{sdn}} \ll \tau_{\tbox{cl}}$ & \ \\
\ & \ \\
At \ \ $t = t_{\tbox{prt}}$ &  
$b(t) \sim s(t) \sim b \ll (\hbar/t)/\Delta$ \\   
At \ \ $t = t_{\tbox{sdn}}$ & 
$b \ll s(t) \sim  (\hbar/t)/\Delta$  \\
At \ \ $t = \tau_{\tbox{cl}}$ & 
$b \sim  (\hbar/t)/\Delta \ll s(t)$ \\
\ & \
\end{tabular}
} \hspace*{1cm} } 
\end{center}
\caption{\protect\rm\footnotesize 
Various time scales in the route to stochastic behavior.}
\end{table}

It is now appropriate to discuss the quantum mechanical 
sudden approximation. For the perturbative scenario 
($v_{_{\tbox{PR}}}\ll 1$) we have already mentioned 
that the AD departs from the PD at $t_{\tbox{sdn}}=\tau_{\tbox{cl}}$, 
which is the time to resolve the energy scale $\Delta_b$. 
In case of the non-perturbative scenario ($v_{_{\tbox{PR}}}\gg 1$)
there is an {\em earlier breakdown} of the 
quantum mechanical sudden approximation. This is 
because we have $\tau_{\tbox{cl}} \gg t_{\tbox{prt}}$ 
and consequently at $t=\tau_{\tbox{cl}}$ we 
already have $s(t)\gg b$. Therefore $t_{\tbox{sdn}}$ should 
be defined as the time to resolve the energy scale 
which is associated with $s(t)$. It leads to
\begin{eqnarray}
t_{\tbox{sdn}} \ = \ b^{\tbox{1/4}} 
(\tau_c^{\tbox{qm}} \tau_{\tbox{cl}})^{\tbox{1/2}}
\ = \ 
\left( \frac{\hbar^2 \tau_{\tbox{cl}}}
{\nu_{\tbox{E}}V^2}\right)^{1/4}
\hspace*{1cm} \mbox{for $v_{\tbox{PR}}\gg 1$}
\end{eqnarray}
The various time scales are summarized in Table II.   
The non-perturbative crossover from genuine-ballistic 
to genuine-diffusive behavior in not trivial. 
If  $v_{_{\tbox{SC}}} \gg 1$ we can relay on semiclassical 
considerations in order to establish the existence 
of this crossover. More generally, for $v_{_{\tbox{PR}}} \gg 1$, 
we would like to have (but we do not have yet) 
an appropriate effective RMT model. 
This effective RMT model should support genuine-ballistic 
motion, and to be further characterized by an 
elastic scattering time~$\tau_{\tbox{cl}}$.

\section{The quantum mechanical adiabatic approximation}
   
The previous analysis has emphasized the 
role of core-to-tail transitions in energy spreading. 
Our assumption was that these transitions are not 
suppressed by recurrences. This is not true  
in the quantum-mechanical adiabatic regime 
($v_{\tbox{LZ}}\ll 1$). Following \cite{wilk} 
it is argued that energy spreading in the 
latter regime is dominated (eventually) by 
Landau-Zener transitions between near-neighbor 
levels. One obtains 
\begin{eqnarray}  
D_{\tbox{E}}^{\tbox{LZ}} \ \approx \ 
\left(\frac{1}{v_{\tbox{LZ}}}\right)^{1{-}(\beta/2)}
D_{\tbox{E}}^{\tbox{cl}} 
\hspace*{2cm} \mbox{for \ $v_{\tbox{LZ}}\ll 1$}       
\end{eqnarray}
where $D_{\tbox{E}}^{\tbox{cl}}$ is the classical result. 
The non-trivial nature of Landau-Zener transitions and the 
statistics of the avoided-crossings is taken into account. 
One should use $\beta=2$ for the Gaussian unitary ensemble (GUE) 
and $\beta=1$ for the Gaussian orthogonal ensemble (GOE).

\newpage 
\section{Classical Brownian motion}

It is possible to argue that the reduced motion of the particle 
(`piston' if the motion is constrained to one dimension) 
obeys the following Langevin equation:
\begin{eqnarray} \label{e81}
m\ddot{\mbf{x}}+\eta\dot{\mbf{x}}={\cal F}
\end{eqnarray}
where $\eta=\mu$ is the friction coefficient.  
The stochastic force is redefined 
\mbox{${\cal F} := {\cal F}{-}\langle {\cal F} \rangle$}, 
such that it is zero on the average. Its 
second moment should satisfy 
\begin{eqnarray} \label{e82}  
\langle{\cal F}(t){\cal F}(t')\rangle_{\tbox{at $x$}} 
\ = \ \phi(t{-}t') 
\end{eqnarray}
where $\phi(t{-}t')=C(t{-}t')$. Usually, it is further 
assumed that higher moments are determined 
by Gaussian statistics. 
The correlation time is denoted as before 
by~$\tau_{\tbox{cl}}$. 
It is better to view the stochastic force  
as arising from a stochastic potential    
\begin{eqnarray}  \label{e83}  
{\cal F}(t) \ \ = \ \ {\cal F}(t,\mbf{x}(t)) 
& \ = \ & -\nabla{\cal U}(\mbf{x},t)    \\  \label{e84}
\langle{\cal U}(\mbf{x}'',t''){\cal U}(\mbf{x}',t')\rangle
& \ = \ & \phi(t''{-}t')\cdot w(\mbf{x}''{-}\mbf{x}') 
\end{eqnarray}
The spatial correlations of the stochastic potential 
are assumed to be characterized by a spatial scale $\ell$. 
The natural tendency is to identify 
$\ell$ with $\delta x_c^{\tbox{cl}}$, but this 
point deserves further non-trivial discussion.  
The normalization convention $w''(0)=-1$ is used, 
and therefore there is consistency of (\ref{e84}) 
with (\ref{e82}). 

\section{The DLD Hamiltonian}

Formally, the Langevin equation (\ref{e81}) with 
(\ref{e84}) is an exact description of the reduced 
dynamics that is generated by the DLD Hamiltonian:
\begin{eqnarray} 
{\cal H} \ = \ \frac{\mbf{p}^2}{2m}  
\ + \ \sum_{\alpha} c_{\alpha} Q_{\alpha} 
u(\mbf{x}{-}\mbf{x}_{\alpha})
\ + \ \sum_{\alpha}\left
(\frac{P_{\alpha}^2}{2m_{\alpha}}
+\frac{1}{2} m \omega_{\alpha}^2 Q_{\alpha}^2\right)
\end{eqnarray}
where $\mbf{x}_{\alpha}$ is the location of the 
$\alpha$ oscillator, $u(\mbf{x}-\mbf{x}_{\alpha})$ 
describes the interaction between the particle 
and the $\alpha$ oscillator, and $c_{\alpha}$ 
are coupling constants. It is assumed that the 
function $u(\mbf{r})$ depends only on $|\mbf{r}|$. 
The range of the interaction is $\ell$. 
The oscillators are distributed uniformly all 
over space. Locally, the distribution of their 
frequencies is ohmic.  Namely, 
\begin{eqnarray}    
\frac{\pi}{2} \sum_{\alpha}
\frac{c^2_{\alpha}}{m_{\alpha}\omega_{\alpha}} 
\delta(\omega-\omega_{\alpha}) \ 
\delta(\mbf{x}-\mbf{x}_{\alpha})
\ = \ \eta\omega  \ \ \ \ 
\mbox{for $\omega<1/\tau_{\tbox{cl}}$} \ \ \ \ .
\end{eqnarray}
This distribution is uniquely determined by the 
requirement $\phi(\tau)=C(\tau)$. The spatial correlations 
are determined via 
\begin{eqnarray} \label{e12}
w(\mbf{r})=\int_{-\infty}^{\infty}
u(\mbf{r}{-}\mbf{x}')u(\mbf{x}')dx'
\end{eqnarray}
For example, we may consider a Gaussian $u(\mbf{r})$ 
for which 
\begin{eqnarray}    \label{e13}
w(\mbf{r}) \ = \ \ell^2\exp
\left(-\frac{1}{2}\left(\frac{\mbf{r}}{\ell}\right)^2\right)  
\end{eqnarray}
Certain generalizations of this assumption has 
been considered in \cite{qbm}, but are of no interest 
here. In the formal limit $\ell\rightarrow\infty$ 
the DLD model reduces to the well known ZCL model. 
The ZCL model is defined by the interaction term 
$x\cdot\sum c_{\alpha} Q_{\alpha}$.

\begin{figure}[h] 
\begin{center}
\leavevmode 
\mbox{\epsfysize=4in \epsffile{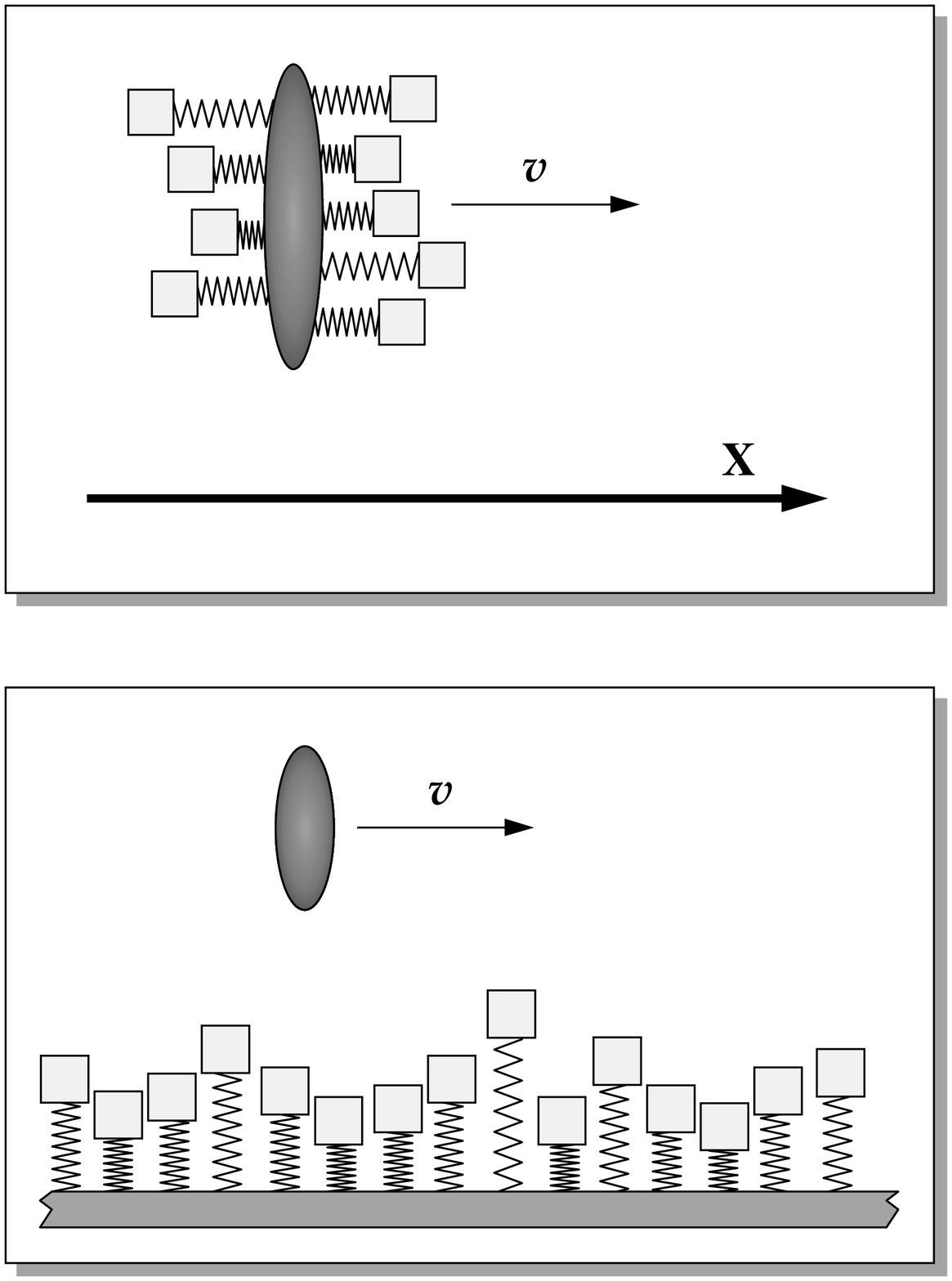}} 
\mpb{3}{
\mbox{\hspace*{0.4cm} 
\epsfxsize=2.0in \epsfysize=1.5in \epsffile{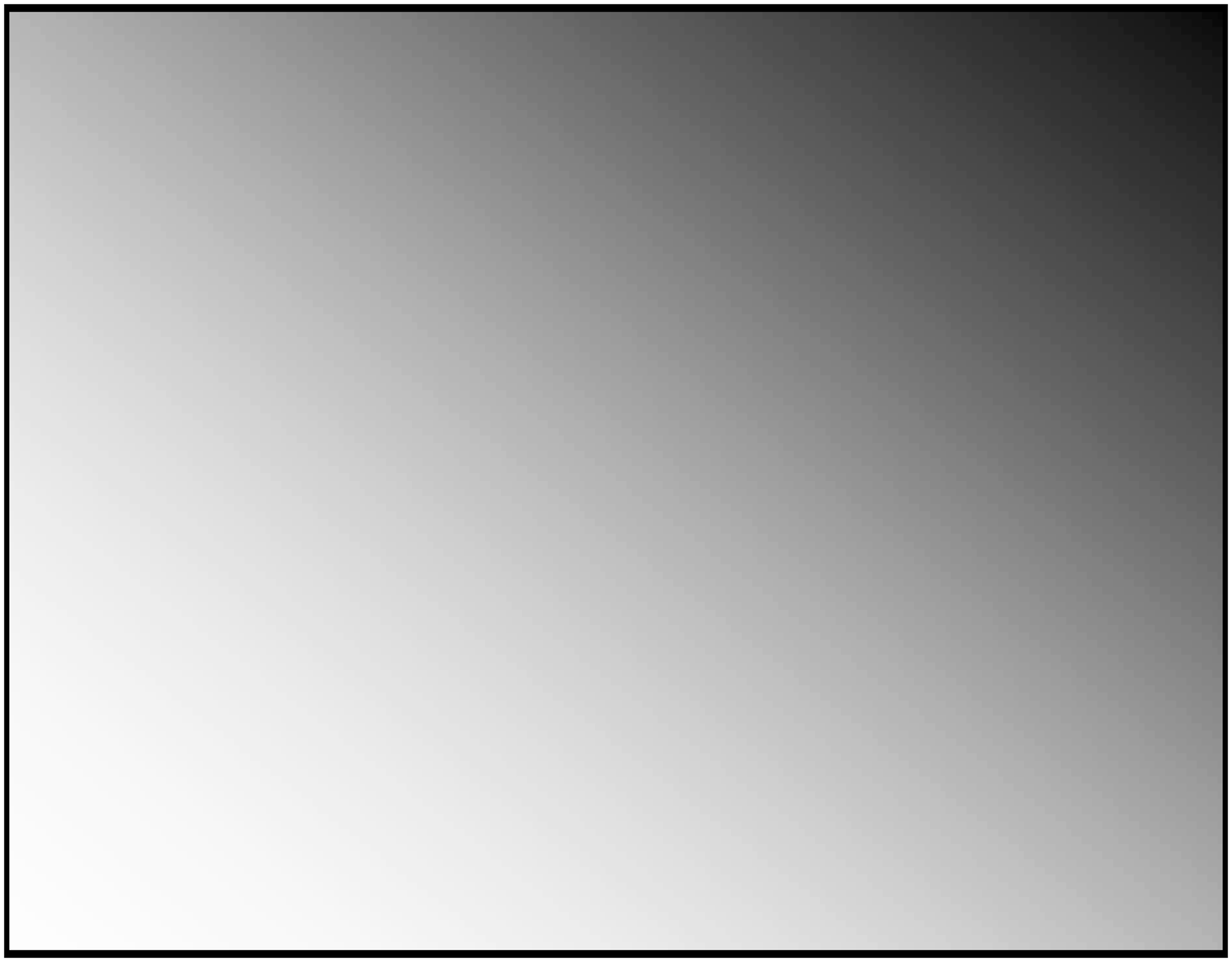}} \\
\vspace*{-0.5cm}
\mbox{\epsfysize=2.15in \epsffile{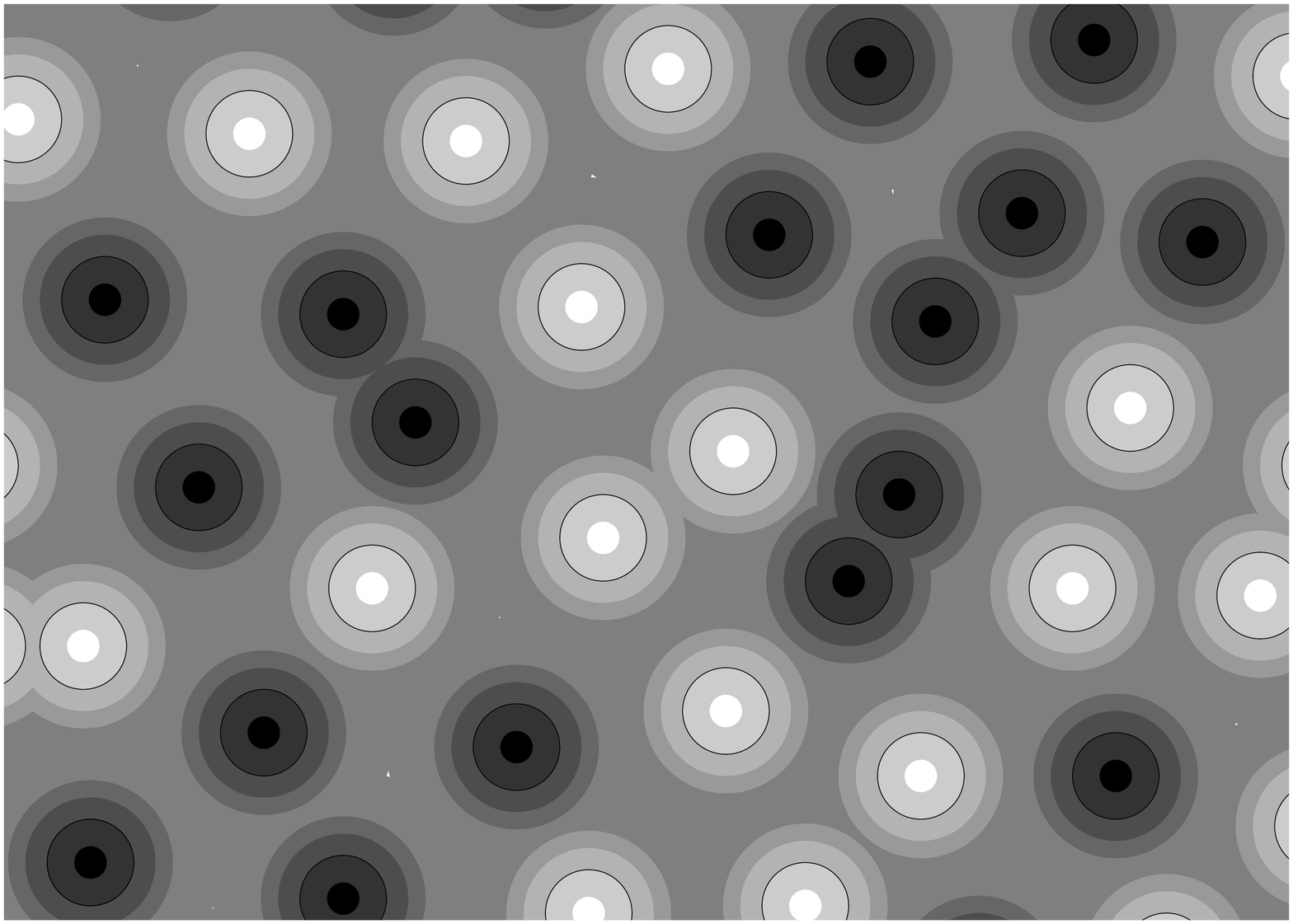}} 
}
\end{center}
\caption{\protect\footnotesize Illustration of the ZCL 
model (left upper drawing), versus the DLD model (left lower drawing). 
The instantaneous potential that is experienced 
by the particle is either linear (right upper drawing), 
or of disordered nature (right lower drawing) correspondingly. 
If the fluctuations are uncorrelated in time (WNA), then the 
two models are classically equivalent. There is no such equivalence 
in the quantum-mechanical case.}
\label{f_models}
\end{figure}

\section{The white noise approximation (WNA)}

One wonders whether the noise in Langevin equation 
can be treated as white noise, meaning that 
$\tau_{\tbox{cl}}$ is irrelevant and we can 
set $\tau_{\tbox{cl}} \sim 0$ in any significant 
result. The following condition defines the 
classical notion of white noise: 
\begin{eqnarray} 
\mbox{Generic Brownian Motion} 
\ \ \ \  \Leftrightarrow \ \ \ \  
\frac{v}{\ell} \ll \frac{1}{\tau_{\tbox{cl}}} 
\end{eqnarray}
If the condition is not satisfied, then 
$\tau_{\tbox{cl}}$ is larger than $\ell/v$, 
and  the particle performs a stochastic  
motion that depends crucially on the 
`topography' of the stochastic potential.  
Note that upon the identification of 
$\ell$ with $\delta x_c^{\tbox{cl}}$ the  
condition above becomes equivalent to the 
trivial requirement of classical slowness (Sec.5).

In the quantum mechanical analysis it is found that 
$\phi(\tau)$ is further characterized by a 
correlation time $\hbar/(k_{\tbox{B}}T)$. Thus, 
the following additional requirement should be met 
if we wish to apply a white noise approximation.
\begin{eqnarray} 
\mbox{High Temperatures} 
\ \ \ \  \Leftrightarrow \ \ \ \  
\frac{v}{\ell} \ll \frac{k_{\tbox{B}} T}{\hbar} 
\end{eqnarray}
Note that the smallest meaningful velocity is the 
thermal velocity, and therefore the above condition 
cannot be satisfied unless the thermal wavelength 
$\lambda_{\tbox{T}}=\hbar/(mk_{\tbox{B}} T)$ is much 
smaller than $\ell$.  From now on we 
assume that both classically and quantum mechanically
we can use the white noise approximation. 
Thus we can use the formal substitution 
$\phi(\tau)=\nu\delta(\tau)$.

\section{Consequences of the WNA}

In the general case,~(\ref{e82}) is less informative than 
(\ref{e84}). However, in case of a {\em classical} particle that 
experiences {\em white noise}, the additional information 
is not required at all! 
Classically, the spatial correlations, and hence $\ell$, are of 
no importance. This is because at each moment a classical 
particle samples {\em one} definite point in space. 
The particle `does not care' about the force elsewhere. 
In the quantum mechanical case the latter observation     
is wrong. A quantum mechanical particle samples each moment 
a {\em finite} region in space and therefore 
spatial correlations of the effective stochastic 
potential become important, even if the noise is white.

Both classically and quantum mechanically the reduced 
dynamics that is generated by the DLD model can be obtained 
analytically and cast into the form 
\begin{eqnarray} 
\rho_t(R,P) \ = \ \int\int dR_0dP_0 \ 
{\cal K}(R,P|R_0,P_0)\rho_{t=0}(R_0,P_0)
\end{eqnarray}
where $\rho_t(R,P)$ represent either the classical-state 
or the quantum-mechanical-state of the particle. In the 
latter case it is a Wigner function. It is a consequence 
of the WNA that the propagator has a Markovian property.  
Namely, it can be written as the composition of smaller 
`time steps'. Consequently $\rho_t(R,P)$ satisfies 
a master equation of the type
\begin{eqnarray} 
\frac{\partial\rho}{\partial t} \ = \ {\cal L} \rho
\end{eqnarray}
The classical version of this equation is known 
as the Fokker-Planck equation. We shall discuss 
shortly its quantum-mechanical version.

\section{The reduced propagator}

The reduced propagator of the DLD model can be obtained \cite{dld}
classically as well as quantum mechanically 
using a path integral technique. In the quantum mechanical 
case it is known as the FV-formalism. The classical version of 
FV formalism can be regarded as a formal solution of 
Langevin equation. It can be obtained without going  
via the DLD Hamiltonian, but then one should assume Gaussian 
statistics.

\begin{figure}[h] 
\begin{center}
\leavevmode 
\epsfysize=3.4in
\epsffile{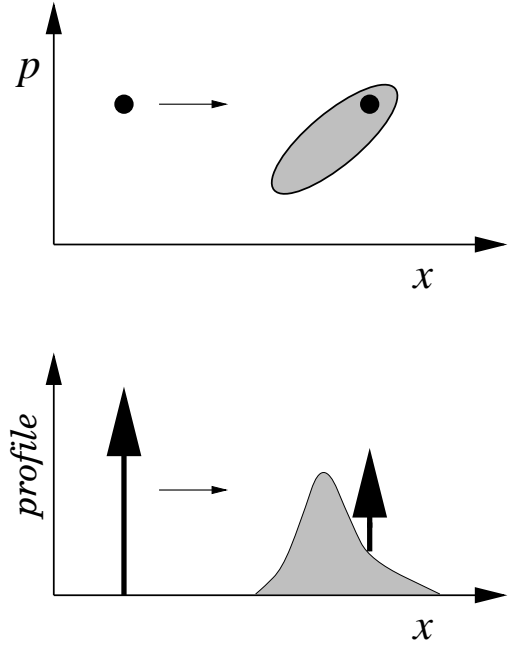}
\end{center}
\caption{\protect\footnotesize 
Upper plot: Phase space illustration for 
the structure of the DLD propagator. 
Lower plot: The projected phase-space density.
Note the existence of an un-scattered component. 
Such component is absent in the ZCL limit as well
as in the classical analysis. } 
\label{f_propg}
\end{figure}

In the absence of coupling to the environment, the 
free-motion propagator of Wigner function is 
the same both classically and quantum mechanically: 
\begin{eqnarray} 
{\cal K}(R,P|R_0,P_0) \ = \ 
{\cal K}_{\mbox{\tiny free}}^{(cl)} 
\ = \ 2\pi\delta(P{-}P_0) \ 
\delta((R{-}R_0)-\mbox{$\frac{P}{m}$}t)
\end{eqnarray}
Taking the environment into account,  
one obtains, in the classical case, 
\begin{eqnarray} \label{e97}
{\cal K}(R,P|R_0,P_0) \ = \ 
{\cal K}_{\mbox{\tiny damped}}^{(cl)} 
\ = \ \mbox{Gaussian} 
\end{eqnarray}
This result is, as a-priori expected,  
independent of $\ell$.  The average velocity of 
of the particle goes to zero on a time 
scale $\tau_{\eta}=(\eta/m)^{-1}$, while the 
spreading of the Gaussian~is  
\begin{eqnarray} 
\delta p^2 \ = & \ \nu t 
\ \ \ & \mbox{for $t\ll\tau_{\eta}$} \\
\delta x^2 \ = & \ (1/(3m^2))\nu t^3  
\ \ \ \ \ \ \ \ & \mbox{for $t\ll\tau_{\eta}$} \\
\delta p^2 \ = & \ mk_{\tbox{B}}T  
\ \ \  & \mbox{for $t\gg\tau_{\eta}$} \\
\delta x^2 \ = & \ (\nu/\eta^2) t 
 \ \ \  & \mbox{for $t\gg\tau_{\eta}$}
\end{eqnarray}

\ \\
The above result holds also in the quantum mechanical 
case provided we take the limit $\ell\rightarrow\infty$. 
No genuine quantum-mechanical effects are found 
if the ZCL model is used to describe Brownian motion! 
Recall again that we are considering here the 
high-temperature case where the WNA applies.  
Now we want to discuss the finite $\ell$ case.  
Here one obtains \cite{dld,qbm} the following expression:
\begin{eqnarray} \label{e99}
{\cal K} \ = \ 
W_{\hbar/\ell}\star {\cal K}_{\mbox{\tiny damped}}^{(cl)}
\ + \ \mbox{e}^{-\frac{2\eta k_BT\ell^2}{\hbar^2}t} 
(1{-}W_{\hbar/\ell}\star) \ {\cal K}_{\mbox{\tiny free}}^{(cl)}
\end{eqnarray}
As suspected, unlike in the classical case, the quantum mechanical 
result depends on $\ell$ in an essential way. 
$W(R{-}R',P{-}P')$ is a smooth Gaussian-like kernel that  
has unit-normalization. Its spread in phase space is 
characterized by the momentum scale $\hbar/\ell$, and 
by an associated spatial scale. The symbol $\star$ stands for 
convolution. Thus, the classical propagator is 
smeared on a phase-space scale that correspond to 
$\Delta p = \hbar/\ell$ and there is an additional 
un-scattered component that decay exponentially and 
eventually disappears. The structure of the propagator 
is illustrated in Fig.\ref{f_propg}. The significance 
of this structure will be discussed shortly. 

\section{Master equation}

To write down an explicit expression for the 
propagator is not very useful. Rather, it is 
more illuminating to write 
an equivalent Master equation. This is 
possible since at high temperatures the 
propagator possess a Markovian property. 
The final result \cite{qbm} is  
\begin{eqnarray} \label{e100}
\frac{\partial\rho}{\partial t} \ = \ \left[
\ - \ \partial_R \ \frac{1}{m} P
\ + \ \partial_P \ \frac{\eta}{m} G_F \star P
\ + \ \nu G_N \star  
\right] \ \rho
\end{eqnarray}
\begin{eqnarray} 
\nonumber 
G_F \ &\equiv \ \ \ \ \ 
{\cal FT} \left(\frac{w'(r)}{r}\right)
\ &= \ \frac{1}{\hbar/\ell}
\hat{G}_F\left(\frac{P{-}P'}{\hbar/\ell}\right)
\\  
\nonumber
G_N \ &\equiv \ \frac{1}{\hbar^2}
{\cal FT} \left( w(r){-}w(0) \right) 
\ &= \ 
\left(\frac{\ell}{\hbar}\right)^2
\left[ \frac{1}{\hbar/\ell}
\hat{G}_N\left(\frac{P{-}P'}{\hbar/\ell}\right) 
- \delta(P{-}P')\right] 
\end{eqnarray}
For convenience the friction kernel $G_F$ and 
the noise kernel $G_N$ are expressed in term 
of smooth Gaussian-like scaling functions
$\hat{G}_F$ and $\hat{G}_N$ that are 
properly normalized to unity. The notation ${\cal FT}$ 
stands for Fourier transform.

If Wigner function does not possess fine 
details on the momentum scale $\hbar/\ell$, 
then the convolution with $G_F$ can 
be replaced by multiplication with $1$, 
and the convolution with $G_N$ can 
be replaced by ${\partial^2}/{\partial P^2}$. 
These replacements are formally legitimate 
both in the classical limit 
$\hbar \rightarrow 0$, and in the ZCL limit 
$\ell \rightarrow \infty$. One obtains 
then the classical Fokker Planck equation   
\begin{eqnarray} \label{e27} 
\frac{\partial\rho}{\partial t} \ = \ \left[
\ - \ \partial_R \ \frac{1}{m} P
\ + \ \partial_P \ \frac{\eta}{m} P
\ + \ \nu \frac{\partial^2}{\partial P^2}  
\right] \ \rho
\end{eqnarray}
The Fokker Planck equation is just a 
continuity equation with an added noise term 
that reflects the effect of the stochastic force.  
The Fokker Planck equation is equivalent to 
solving Langevin equation as well as to the 
Gaussian propagator~(\ref{e97}).

\section{Brownian motion and dephasing}

Wigner function may have some modulation on 
a fine scale due to an interference effect. 
The standard text-book example of a two slit 
experiment is analyzed in \cite{qbm}. 
In case of the ZCL model ($\ell=\infty$), the propagator 
is the same as the classical one, and therefore 
we may adopt a simple Langevin picture in 
order to analyze the dephasing process.      
Alternatively, we may regard the dephasing 
process as arising from Gaussian smearing of 
the interference pattern by the propagator.   
In case of the DLD model (finite~$\ell$) we should 
distinguish between two possible mechanisms 
for dephasing: 
\begin{itemize}
\item Scattering (Perturbative) Mechanism. 
\item Spreading (Non-Perturbative) Mechanism.
\end{itemize}
Actually, it is better to regard  
them as mechanisms to maintain coherence. 
The first mechanism to maintain coherence 
is simply not to be scattered  
by the environment. The second mechanism 
to maintain coherence is not to be 
smeared by the propagator. The first  
mechanism is absent in case of the   
ZCL model. 

Let us discuss how coherence is lost due to  
the scattering mechanism.  The discussion is 
relevant if Wigner function contains  
a modulation on a momentum scale 
much finer than $\hbar/\ell$, else $\ell$ becomes 
non-relevant and we can take it to be infinite. 
One should observe that such modulation is not 
affected by the friction, but its intensity decays 
exponentially in time. This is based 
on inspection of either the propagator~(\ref{e99}), 
or the equivalent Master equation~(\ref{e100}). 
In the latter case note that 
the convolution with $G_N$ can be replaced 
by multiplication with $-(\ell/\hbar)^2$. 
The decay rate is     
\begin{eqnarray} 
\frac{1}{\tau_{\varphi}}  
\ = \ \frac{2\eta k_BT\ell^2}{\hbar^2} 
\ \ \ \ \ \ \ \ \ \mbox{assuming WNA} 
\end{eqnarray}
This is the universal result for the dephasing 
rate due to the `scattering mechanism'. 
It is universal since it does not depend on 
details of the quantum-mechanical state involved. 
However, the validity of this result is  
restricted to the high temperature 
regime, where the WNA can be applied. Extensions 
of this result, as well as discussion of  
dephasing at low temperatures can be found 
in \cite{qbm,dph}. 
 
\section{The open question}

The effective-bath approach suggests that there 
is a universal description of quantal Brownian motion. 
If indeed the effective-bath approach is universally 
applicable, it is implied that 
\begin{eqnarray} \label{e_open}
\mbox{chaos} \hspace*{1cm} 
\Longleftrightarrow \hspace*{1cm} 
\mbox{dynamical disorder}
\end{eqnarray}
meaning that the motion under the influence of 
chaotic environment is effectively the same as 
the motion under the influence of dynamical 
disordered environment.

At this stage the conditions for the validity 
of (\ref{e_open}) are yet unclear. It is interesting 
however to emphasize the practical implications 
of such claim. The starting point should be 
a specification of the effective correlation 
scale $\ell$. For the `piston' example the most 
obvious {\em guess} is 
\begin{eqnarray} \label{e_guess}
\ell \ = \ \left\{ \matrix{
\delta x_c^{\tbox{cl}} & 
\ \ \ \mbox{if} \gg \lambda_{\tbox{E}}  \cr 
\lambda_{\tbox{E}} & \ \ \ \mbox{\small else} } \right.
\end{eqnarray}
One may estimate now the coherence time using the 
substitution $\nu= \mathsf{m}^2 v_{\tbox{E}}^3 / L$ where 
$\mathsf{m}$ is the mass of the gas particle, and 
$L=v_{\tbox{E}}\tau_{\tbox{col}}$ is the mean 
path-length between collisions with the piston.
The result is 
\begin{eqnarray} 
\tau_{\varphi} \ \ = \ \ \frac{\hbar^2}{\nu \ \ell^2} 
\ \ = \ \ \left(\frac{\lambda_{\tbox{E}}}{\ell}\right)^2 \tau_{\tbox{col}} 
\ \ \le \ \ \tau_{\tbox{col}}
\end{eqnarray}
where $\lambda_{\tbox{E}}$ is the De-Broglie 
wavelength of the gas particle. Note that we 
are assuming the WNA, and therefore we must 
satisfy $\lambda_{\tbox{B}}\ll \ell$, where 
$\lambda_{\tbox{B}}$ is the De-Broglie 
wavelength of the piston.

The (effectively) disordered nature of the  
environment is significant only within the 
time domain $t<\tau_{\varphi}$. The non-trivial 
effect that is `predicted' by solving the DLD model 
is that the reduced propagator has a  
coherent `unscattered-component' plus a smearing  
'scattered-component'. The latter is created due to the exchange 
of momentum-quanta of typical magnitude $\hbar/\ell$. 
Assuming `hard walls' (in the sense of (\ref{e_guess})) 
we get the result $\hbar/\ell \sim 2\mathsf{m}v_{\tbox{E}}$ 
and $\tau_{\varphi}=\tau_{\tbox{col}}$. 
This is a very 'funny' result since it has a 
trivial {\em classical interpretation} in terms of    
the actual ('piston') model, whereas within the 
effective-bath approach it appears as a 
genuine quantum-mechanical effect! 
 
\acknowledgments
I thank Uzy Smilansky and Eric Heller for interesting  
discussions, and Shmuel Fishman for fruitful interaction 
in intermediate stages of the study.   

\newpage

\end{document}